\documentclass[journal]{vgtc}                     

\onlineid{1615}

\vgtccategory{Research}

\title{\textit{Skeleton}: Visual Authoring of Non-visual Data Experiences}

\author{%
  \authororcid{Frank Elavsky}{0000-0002-6849-5893},
  Chieri Nnadozie,
  \authororcid{Lucas Nadolskis}{0009-0002-1079-2518},
  \authororcid{Patrick Carrington}{0000-0001-8923-0803} and 
  \authororcid{Dominik Moritz}{0000-0002-3110-1053}
}

\authorfooter{
    \item Elavsky, Nnadozie, Carrington, and Moritz are with Carnegie Mellon University. E-mail: \{ fje | pcarrington  | domoritz \}@cmu.edu, cnnadozi@andrew.cmu.edu
    \item Nadolskis is with UC Santa Barbara. E-mail: lgilnadolskis@ucsb.edu
}

\abstract{%
  When sighted practitioners author accessible data visualizations, they build navigation structures (the nodes, edges, and input bindings that govern how assistive technologies traverse an interface) entirely in code, with no visual representation. This invisibility makes navigation structures difficult to inspect, debug, and iterate on. To sighted practitioners, every other aspect of a visualization is iterated on because it is visible; navigation structure ships as a first draft, if at all, because it is not. Without a representation to react to, practitioners cannot develop judgment about what makes navigation good or bad, and the quality ceiling of non-visual experiences is set by the absence of a feedback loop. We address this problem through longitudinal co-design with practitioners across cartography, design systems, and open-source visualization, and make three contributions. First, we introduce technical advancements for making the properties of accessible navigation structure visible and directly manipulable during authoring, grounded in two foundational pieces of infrastructure produced by our co-design work: an \textit{Inspector} that renders navigation graphs as interactive node-link diagrams, and a \textit{Dimensions API} that expresses navigation in terms of data dimensions rather than explicit graph construction. Second, building on these, we present \textit{Skeleton}, a direct-manipulation authoring environment in which the properties of an accessible navigation structure are translated into visual representations authors can observe and manipulate. Key techniques include a dual-view editor that simultaneously shows the system's navigation model and the end user's spatial experience, a scaffolding engine that automates spatial node placement by repurposing a visualization rendering pipeline, a live label-template editor with real-time screen-reader-output preview, and a testing mode that makes traversal sequence visually trackable. Third, we evaluate \textit{Skeleton} through an in-situ study with 8 practitioners across visualization design, engineering, and research. Making navigation structure visible changed how practitioners engaged with accessible design: they reconsidered the architecture of their own visualizations, attended to a broader range of input modalities, and shifted from treating accessibility as a compliance task to treating it as a design problem.

  A free copy of this paper and all supplemental materials are available at \url{https://osf.io/7afw4}.
}


\teaser{
  \centering
  \includegraphics[width=\linewidth, alt={Skeleton's interface. There is a tree-like structure with 3 hierarchical levels shown in the schema. In the editor, outlines shown over the 3 lines in a line chart, along with vertical bands that join the lines at each month they cross. These two shapes appear like a triple helix. On the right is a panel labeled scaffold, which has options for setting various parameters.}]{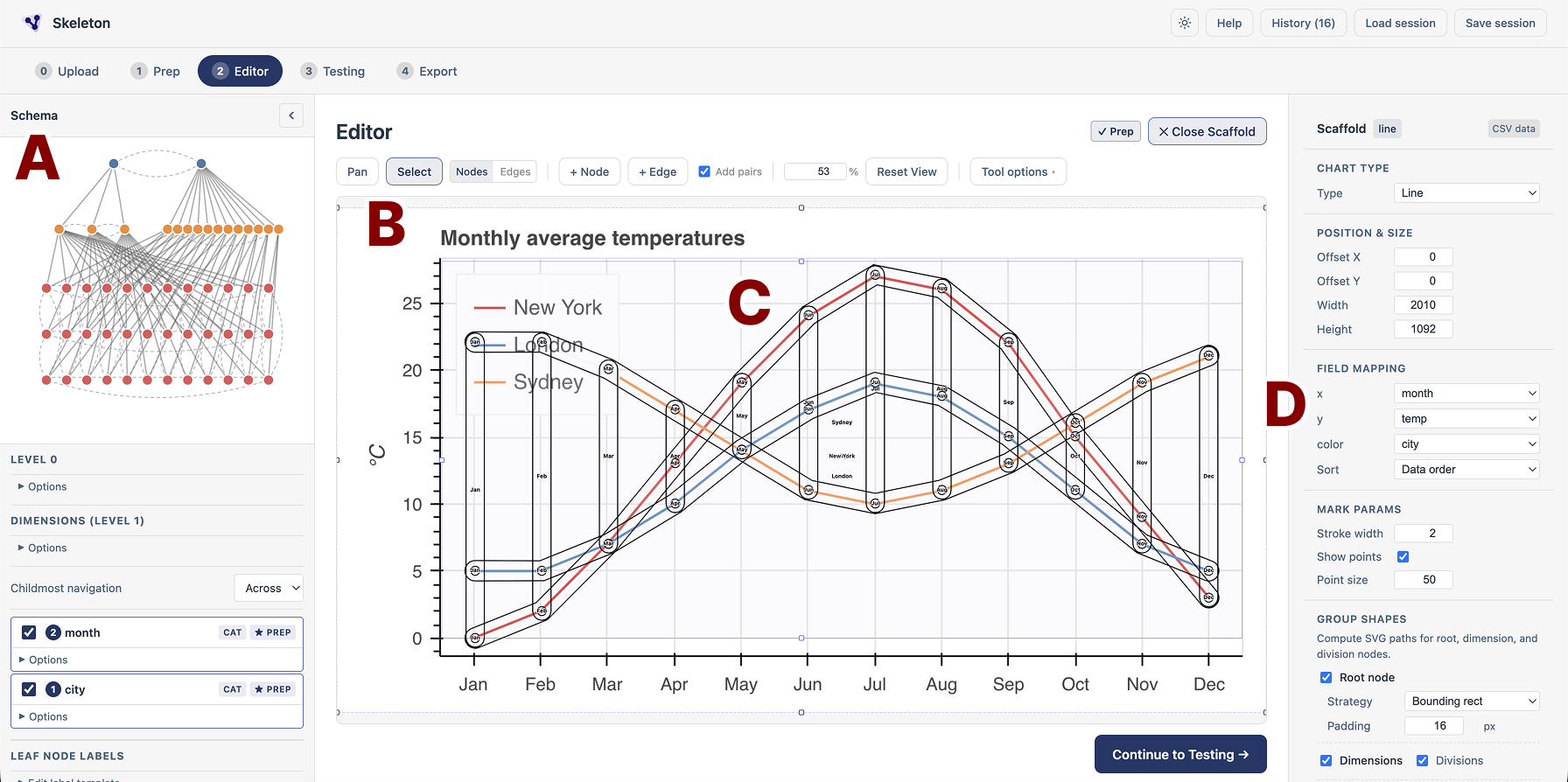}
  \caption{%
  	\textit{Skeleton} being used to build a navigation structure over a line chart. \textbf{A}: a dual tree visualization of nodes and edges built for a line chart with 3 lines, spanning over 12 months. \textbf{B}: a direct-manipulation canvas that resides over a static png image of a line chart. \textbf{C}: a skeleton-like structure of nodes, edges, and group outlines that correspond to the schema shown in (A), overlaid on the visual space of the chart in (B). \textbf{D}: controls for rapidly scaffolding the nodes, edges, and their shapes shown in (C), using Vega's visualization engine for automatically generating spatial properties. %
  }
  \label{fig:hero}
}

\graphicspath{{figs/}{figures/}{pictures/}{images/}{./}}

\usepackage{tabu}
\usepackage{booktabs}
\usepackage{lipsum}
\usepackage{mwe}
\usepackage{ccicons}
\usepackage{mathptmx}

\begin{document}

\firstsection{Introduction}

\maketitle

We start this work with a provocation: How might the discipline of visualization help the discipline of accessibility?

Visualization has spent decades developing techniques for a specific class of problem: representing and interacting with information visually. Grammars of visual encoding~\cite{Satyanarayan2017VegaLite,Wilkinson2011,McNutt2022,ZongAnimated2023,Wickham2010}, direct-interaction interfaces~\cite{Hutchins1985DirectManip,Endert2013,Shneiderman1992}, and iterative feedback loops between representation and understanding~\cite{Heer2007,Fisher2012,Liu2014} are all methods that enable abstraction and manipulation of the information that underlies the visual representation. We argue these methods have a direct application within the discipline's own accessibility challenges.

Sighted practitioners who build accessible data visualizations face an unusual authoring problem. The non-visual navigation structures they construct (the nodes, edges, focus states, input bindings, and semantics that govern how assistive technologies traverse a chart) exist only as code. A practitioner can write a navigation hierarchy, but cannot see it, click on a node to inspect what will be announced, observe the spatial relationship between a navigation path and the chart it overlays, and then manipulate its properties through direct interaction. Every other aspect of a visualization has a visible, inspectable representation during authoring: the visual encodings are visible, the layout is visible, the interaction states are visible. And yet, navigation structure is not.

This invisibility has practical consequences. Without a way to see what they are building, sighted practitioners cannot easily catch structural errors, compare design alternatives, and iterate. Accessibility becomes downstream of every other design choice because the authoring conditions do not support anything else~\cite{Joyner2022Wild, Sharif2024Challenges}. The floor and ceiling of non-visual data experiences are constrained by what sighted authors can perceive of their own work.

We engage this space with the following research questions:
\begin{enumerate}[label=\textbf{R\arabic*}, leftmargin=0pt, itemindent=!, itemsep=0pt, topsep=0.5em, parsep=0pt]
 
\item \label{rq:R1}\textbf{(Qualitative, Exploratory):} What challenges do sighted practitioners face when designing and engineering navigation structures for accessible visualizations?
 
\item \label{rq:R2}\textbf{(Qualitative, Exploratory):} How do sighted authors reason about the non-visual experiences that accompany their visualizations?
 
\item \label{rq:R3}\textbf{(System, Design):} How can we make the properties of accessible navigation structure visible and directly manipulable during authoring?
 
\item \label{rq:R4}\textbf{(Qualitative):} In what ways does a directly manipulable visual representation of navigation structure change how practitioners find errors and improve upon their designs?
 
\end{enumerate}
 
We address these questions through longitudinal co-design with practitioners across cartography, design systems, and open-source visualization, following an action research orientation~\cite{Hayes2011ActionResearch} in which the research team was embedded in each community's active development work. This paper makes three contributions:
 
First, \textbf{we introduce technical advancements} for making the properties of accessible navigation structure visible and directly manipulable during authoring. These techniques are grounded in our co-design collaborations, which produced two foundational pieces of infrastructure: an \textit{Inspector} that renders any navigation graph as an interactive node-link diagram, and a \textit{Dimensions API} that formalizes a declarative grammar for expressing navigation in terms of data dimensions rather than explicit graph construction~(\ref{rq:R1}, \ref{rq:R2}, \ref{rq:R3}).
 
Second, building on this infrastructure, \textbf{we present \textit{Skeleton}, a direct-manipulation authoring environment} in which the topology, spatial mapping, semantics, and input logic of an accessible navigation structure are made visible and directly manipulable. \textit{Skeleton} is built on Data Navigator~\cite{Elavsky2023}, a code-based library for constructing interactive data navigation structures~(\ref{rq:R3}). Our intention with \textit{Skeleton} is to continue to develop it towards a usable, practical system beyond the scope of this research.
 
Third, \textbf{we contribute findings from an in-situ interview study} with 8 visualization practitioners. We qualitatively evaluate Skeleton as a design probe~\cite{Hammad2024GameAware,Hutchinson2003TechnologyProbes}, focusing on how it influences practitioner engagement. We find that making navigation structure visible shifted how participants engaged with accessible design: they reconsidered the architecture of their own visualizations, attended to a broader range of input modalities, and shifted from treating accessibility as a compliance task to treating it as a design problem~(\ref{rq:R1}, \ref{rq:R2}, \ref{rq:R4}).

\section{Related Work}

\subsection{Visualizing Non-visual Structure}

Long before accessibility, computing established a practice of taking structures that are not inherently visual and giving them a visual form that can be seen and acted upon. Sutherland's Sketchpad was one of the earliest; it turned the constraints and topology of a drawing into a manipulable graphical object~\cite{Sutherland1963Sketchpad}. Later, visual programming environments and syntax-directed editors rendered computation and structure as editable visual objects~\cite{Smith1977Pygmalion,Teitelbaum1981Cornell}. Myers later named the distinction this lineage had been drawing, separating \textit{visual programming} from \textit{program visualization}, depicting otherwise non-visual programs graphically~\cite{Myers1990Taxonomies}. As a consequence, many structures we now treat as inherently visual are abstractions whose visual rendering has become so dominant that the representation now stands in for the structure.

Direct manipulation interfaces, in turn, provide a continuous representation of the objects of interest and give rapid, incremental feedback, letting people work directly with a structure they would otherwise only manipulate indirectly~\cite{Shneiderman1983DirectManip, Hutchins1985DirectManip}; grammars of interactive graphics extend the same move to authoring~\cite{Satyanarayan2017VegaLite, Bostock2011D3}. Yet across this tradition, one structure has remained conspicuously without a visual form to manipulate: the UI structure that a screen-reader traverses. This lack of visual representation is especially surprising because screen readers evolved in parallel with the emergence of graphical interfaces~\cite{Thatcher1994}.

\subsection{Non-visual Data Experiences}

Blind people who rely on assistive technologies interact with data in fundamentally different ways than sighted users who use a direct pointer, like a mouse~\cite{Marriott, Kim2021Accessible, Sharif2021}. A substantial body of research has documented what these experiences look like across modalities, and what it takes to make them meaningful. In the context of ``data experiences,'' this paper focuses specifically on interactive navigation structures, but we briefly survey adjacent modalities to situate our contribution.

\textbf{Alternative text and natural language.} A dominant strand of this work concerns the generation and evaluation of textual descriptions of visualizations~\cite{Lundgard2022Accessible, Jung2022Communicating, Kim2023Beyond, Kim2023Explain}. More recent LLM-driven systems and Q/A approaches can caption charts with varying degrees of semantic depth, some at a risk of producing bias~\cite{Farahani2023, Elavsky2025, Kim2023Exploring}. While alt text makes a visualization's message available without sight, it is by nature static: a description conveys what a visualization says, but not how a user might explore or interact with it.

\textbf{Sonification, haptics, and tactile rendering.} Non-visual data experiences extend well beyond text. Sonification encodes data as sound~\cite{Hermann2011, Mansur1985Sound, Brewster2002Visualization}, with declarative grammars emerging for authoring these experiences~\cite{Kim2024}. Haptic and tactile representations offer another channel through refreshable displays, 3D-printed graphics, and multimodal touchscreen interactions~\cite{Butler2021Technology, Marriott2026, Holloway2024, Reinders2025}. Recent systems integrate multiple non-visual representations of the same data~\cite{Seo2024, Holloway2022, Chundury2022Towards}.


\textbf{Interactive navigation structures.} The primary focus of our work centers on the state of research related to structured navigation: the traversal of data points, groupings, and interface elements through assistive technologies and keyboard input~\cite{Sorge2016Polyfilling, WAI2017Keyboard, Zong2022Rich, Elavsky2023}. Existing systems and interfaces have demonstrated that going beyond static descriptions to support hierarchical, traversable data structures meaningfully improves how blind users can explore and reason about charts~\cite{Zong2022Rich, Thompson2023Chart}. Giving users control over the textual tokens surfaced during navigation improves comprehension and agency~\cite{Jones2024}. And more recent work has found that perceptually congruent navigation structures for charts and diagrams can improve goal-driven exploration~\cite{Mei2025}.




\subsection{Authoring Non-visual Data Experiences}

Accessible visualization has historically centered the experiences of disabled users, but a parallel body of work examines the experiences of visualization practitioners working on accessibility.

\textbf{Practitioner challenges.} Research consistently finds that sighted visualization practitioners struggle with accessibility~\cite{Joyner2022Wild, Fan2023Accessibility, Sharif2024Challenges}. Most visualizations in the wild are inaccessible and designers themselves report lacking guidance, especially for complex and interactive graphics~\cite{Joyner2022Wild,Sharif2024Challenges}. And screen reader users experience the downstream effects of these gaps: inconsistent structure, poor keyboard support, and information that is present visually but absent in the accessibility tree~\cite{Sharif2021, Fan2023Accessibility}. The pattern across this work is consistent: the practitioners who build visualizations lack the tools and feedback mechanisms to make non-visual experiences effective, useful, and good.

Across practitioner-centered literature, a recurring finding is that non-visual experiences are treated as downstream of visual design choices, added after the visual representation is finished rather than designed in parallel~\cite{Joyner2022Wild, Lundgard2019Sociotechnical, Sharif2024Challenges, Zong2024}. This sequencing has consequences: what is navigable and how it is structured is constrained by whatever visual decisions came first.


\textbf{Authoring-oriented systems.} A distinct line of work has focused on building authoring tools and libraries that give practitioners more tractable paths to accessible output. Few visualization tools support the kinds of interactive navigation structures that assistive technology users most benefit from~\cite{Kim2023Beyond}. Of those that do, most rely on code-based approaches~\cite{Blanco2022olli,Sharif2018evoGraphs,Sharif2022VoxLens,Elavsky2023}. \textit{Umwelt}~\cite{Zong2024} takes a different and notable approach: it is a structured editing environment where authors specify representations \textit{across} modalities (sonification and visualization) in an integrated interface, where navigation is made available over the visualization using \textit{Olli}~\cite{Blanco2022olli}. The latest work in this space is \textit{Arboretum}, a tool that provides automatic conversion of diagrams to a tabular structure, navigable structure, and tactile representation~\cite{Wimer2026}.

\textbf{Communicating visually, authoring invisibly.} There is a revealing irony across this body of related work: research about navigation structures almost invariably communicates those structures visually. Papers such as ``rich screen reader experiences''~\cite{Zong2022Rich}, \textit{ChartReader}~\cite{Thompson2023Chart}, \textit{Benthic}~\cite{Mei2025}, and \textit{Data Navigator}~\cite{Elavsky2023} each use visual node-link diagrams and hierarchical schematics to explain navigation paths to their readers. The same pattern holds in adjacent domains that involve structuring data for navigation, such as PDF remediation~\cite{Mowar2026iTagPDF}.

Yet, most other non-visual modalities are not only communicated but \textit{authored} in visual and non-visual modalities. Sonification is designed in editors that are both visual and auditory~\cite{Kim2024}. Tactile graphics are composed by transcribers on a graphical canvas before being physically embossed or printed~\cite{Bornschein2015Collaborative}.

\textbf{Inspecting accessibility structure.} In authoring-oriented systems, none provide a visual interface through which practitioners can interactively inspect and manipulate navigation structures as a first-class design material. Structure output is either downstream of code or static visuals. Structure is always \textit{derived} or \textit{specified}; indirectly manipulated. Additionally, verification of the structure across all of these systems requires developers to manually navigate using a screen reader after the structure has been authored and rendered, before returning to the upstream visual design space or code. Navigation structure is thus understood and communicated visually by sighted researchers and designers, yet built entirely without visual feedback by developers; this is the gap we address.

\section{Co-design Foundations}
\label{sec:codesign}


Our research follows an \textit{action research} orientation~\cite{Hayes2011ActionResearch}, in which knowledge is generated by engaging with a community to solve a real problem in-situ, alongside them rather than studying it from the outside. These collaborations started from a shared motivation: practitioners needed to make their existing systems accessible to navigational assistive technologies, using Data Navigator~\cite{Elavsky2023} as a foundation. Across three projects, we worked with 12 individuals outside our research team. Three blind co-designers shaped the work throughout: CD1 and CD2 (anonymous) and a co-author on this paper, Nadolskis. CD1's contributions are noted in \Cref{sec:geomap}, CD2's are noted in \Cref{sec:testing}. Nadolskis contributed to early ideation, problem formation, and framing for the project as a whole, helping define the authoring challenges that \textit{Skeleton} addresses, in addition to feedback on study design.

The co-design literature on accessibility has centered people with disabilities as design partners~\cite{Cullen2019Co, Race2023, deGreef2021Interdependent, Thompson2023Chart, Zong2022Rich, Zong2025}. Our co-design takes also sighted practitioners as partners because the authoring challenges we address happen on their side of the process: it is sighted authors who cannot see what they are building. This work was exempted by our IRB (2024\_00000040).

Two themes converged across all three engagements, and we present them here before describing the individual projects that surfaced them. First, \textbf{practitioners communicated and reasoned about accessible navigation using visual representations}: while designing for language, sound, and structure, our collaborators drew on paper, built wireframes of nodes and edges, and reflected on the design space using visual artifacts. Even when collaborating with blind co-designers, a visual medium was the first language of sighted authors. Second, \textbf{development that followed visual design work faced severe iteration barriers}: verifying a navigation experience required building a working code prototype and manually navigating it with a screen reader. The gap between visual design and code-based scaffolding with manual testing produced repeated mistakes and misinterpretations.

These themes did not arrive as abstractions; each project below surfaced them as concrete friction, and together the three engagements produced five design goals that \textit{Skeleton} was built to satisfy. We state them here and, in the subsections that follow, mark where each emerged:

\begin{enumerate}[label=\textbf{DG\arabic*}, leftmargin=0pt, itemindent=!, itemsep=0pt, topsep=0.5em, parsep=0pt]
\item \label{dg:visible}\textbf{Make structure visible and inspectable.} Render the navigation structure as a manipulable object during authoring, not only as code, so that practitioners can perceive and react to it.
\item \label{dg:abstraction}\textbf{Specify navigation at the level of data.} Provide a higher-level, data-driven abstraction so practitioners describe traversal in terms of data dimensions rather than hand-wiring every node and edge.
\item \label{dg:loop}\textbf{Shorten the iteration loop.} Let practitioners verify a design without a full code build or a manual screen-reader pass.
\item \label{dg:bespoke}\textbf{Support image-only and bespoke visualizations.} Do not require a single underlying dataset, format, or recognizable chart type, since the cases that most need accessible tooling often have neither.
\item \label{dg:input}\textbf{Reach beyond keyboard input.} Accommodate input modalities other than assuming keyboard navigation, including touch and text.
\end{enumerate}

\subsection{Geologic Map}
\label{sec:geomap}

\begin{figure}[h]
  \centering 
  \includegraphics[width=\columnwidth, alt={A zoomed out view of the state of Wisconsin, encoded with a tapestry of dozens of colors and textures. Arrows annotate direction from the legend to map area, a separate area has messy, approximated graph structures laid out, a zoomed in version of a chart inside the graphic has schematics next to it titled "grid navigation." An area in the upper corner demonstrates 2 different styles of drawing tooltips during navigation.}]{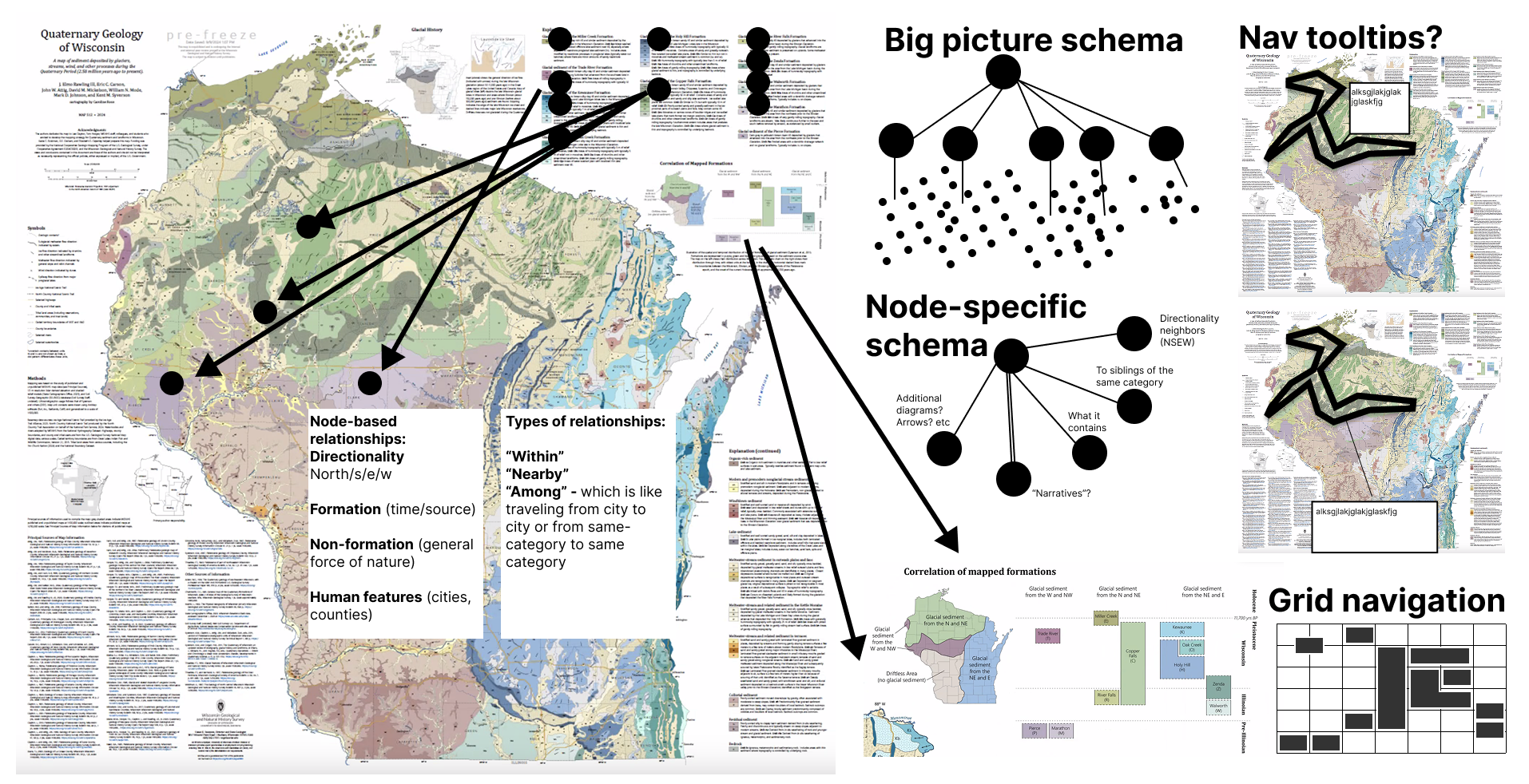}
  \caption{%
  	Our visual design work in Figma over a static geologic infographic map of Wisconsin. We use visual forms and illustrations over and beside the map to communicate flows, structure, navigation styles, and interaction patterns.%
  }
  \label{fig:geo}
\end{figure}

Our longest engagement spanned just over two years, with a cartographer, accessibility consultant, and blind co-designer (CD1) building an interactive version of a quaternary geologic map of Wisconsin~\cite{Rawling2025}. The map combined dozens of irregular geographic regions organized categorically by a legend.

Early design sessions took place in Figma (\Cref{fig:geo}), where we laid out node-link diagrams of how a screen reader user might traverse the map, legend, and peripheral information. We annotated each node with text a screen reader would announce and connected them to relevant regions. Drawing the structure made it possible to discuss design choices, catch dead ends, and debate traversal strategies. We were informally doing what \textit{Skeleton} later formalized.

The friction began when we moved from design to implementation. We had no way to verify that our designed structure would be good or ideal, and scaffolding the project into Data Navigator was arduous: the translation from even simple navigational designs to functional code was too complex to hand off or meaningfully iterate on. The collaboration also surfaced a design-space boundary: for within-map spatial navigation, an egocentric audio-game approach~\cite{Biggs2022Audiom} fit better than a node-link graph, a distinction CD1 helped surface. Reaching this boundary was only possible because we could see the structure we were trying to build (\ref{dg:visible}). And the map's lack of a single tabular dataset made clear that a useful navigation tool should not assume structured data or a grammar of graphics (\ref{dg:bespoke}).

\subsection{Design System Library}
\label{sec:designsystem}

Our second engagement, spanning 7 months, was with 6 engineers and designers on Adobe's React Spectrum Charts library, an open-source chart component system. We worked in their codebase and in Miro on navigation design for bar charts, clustered and stacked bars, line charts, and related types. Because we needed generalized, reusable patterns, our design problems differed from the geologic map: we needed to account for use, re-use, and edge cases across chart types.

Our Miro sessions produced two kinds of artifacts: diagrams of navigation structure for specific chart \textit{instances} and \textit{schema} diagrams capturing generalized patterns in dimensional terms. The concept of a ``dimension'' emerged naturally: common transformations on Data Navigator's graph structure corresponded to properties within a dataset, such as a ``categorical'' dimension or a ``numerical'' dimension, where traversal took place within collections of grouped siblings. This dimensional thinking directly foreshadowed the Dimensions API (\Cref{sec:infrastructure}).

The dominant friction was iteration speed. We could sketch and converge on a schema in Miro in an hour, but verifying the design in a functional example required embedding Data Navigator into a large codebase, implementing changes, rebuilding, and manually testing with a screen reader. The collaboration also surfaced a limitation: mobile screen reader navigation uses swipe gestures rather than keyboard input, and our keyboard interaction model did not account for it. Each of these frictions pointed somewhere: the dimension concept toward a higher-level abstraction (\ref{dg:abstraction}), the cost of verifying each change toward a faster loop that does not require a full build or screen reader (\ref{dg:loop}), and the mobile-gesture gap beyond keyboard input (\ref{dg:input}).

\subsection{Open Source Visualization Library}
\label{sec:opensource}

Our third collaboration, spanning nearly two years, was with Quansight Labs and contributors to Bokeh, a Python-based open-source visualization library. We performed an accessibility audit~\cite{Eswaramoorthy2025BokehAudit} based on Chartability~\cite{Elavsky2022Chartability} and identified that Bokeh visualizations with interactive chart elements needed to be navigable by assistive technologies.

Unlike Adobe, Bokeh has no native chart types. Its API operates at the level of glyphs, renderers, and data sources, which users assemble freely. There was no standard unit around which to anchor a navigation pattern, and any grammar or tooling would need to accommodate an open-ended range of encoding combinations. The iteration gap from Adobe was present in a more severe form: making library-wide contributions to a fully open-source project required incremental tooling for the most common cases. For Bokeh, we needed to test functional, data-driven navigation abstractions without fully embedding Data Navigator into the library. A library with no native chart types required the abstraction to be data-driven and chart-type-agnostic rather than tied to fixed templates, sharpening \ref{dg:abstraction} into a concrete constraint.

\subsection{Infrastructure from Practice}
\label{sec:infrastructure}

The three collaborations converged on the design goals above. Two of them demanded new infrastructure before \textit{Skeleton} could exist: a way to visually render and inspect navigation structures (\ref{dg:visible}), and a higher-level, data-driven abstraction for specifying navigation without hand-wiring every node and edge (\ref{dg:abstraction}). We built two pieces of infrastructure to address these, described next; the remaining goals (\ref{dg:loop}, \ref{dg:bespoke}, \ref{dg:input}) are met by \textit{Skeleton}'s authoring environment (\Cref{sec:skeleton}). These infrastructural improvements as well as \textit{Skeleton} are open source on \href{https://github.com/cmudig/data-navigator}{our GitHub repo}.

\subsubsection{An Inspector Gadget}

We built an \textit{Inspector} (\texttt{@data-navigator/inspector}) to render any Data Navigator structure as an interactive node-link graph using D3, with an accompanying console for debugging. Hierarchical structures are colored by level; edges are drawn as directed links; the entry point is visually marked. The \textit{Inspector}'s graph can itself be navigated using Data Navigator, with visual focus tracking during navigation, allowing practitioners to manually verify structure and reachability.

This made structural verification immediate: a practitioner could generate a structure and check at a glance whether the hierarchy had the right levels, whether circular extents produced expected wrap-around edges, or whether a particular path was reachable. The interactive console logs API information and underlying data when nodes are activated, and hovering or focusing logged information highlights the corresponding node in the graph.

The \textit{Inspector} remains a developer tool, however. It requires code familiarity to attach and renders structure as an abstract graph with no connection to the spatial layout of the underlying visualization. It shows topology but not instantiated geometry: a practitioner can see that two nodes are connected but not where their focus indicators will appear on-screen. This gap between navigable structure and its spatial instantiation over a rendered chart motivated \textit{Skeleton}.

\subsubsection{Alternative Dimensions}

The original Data Navigator library requires explicit graph construction: practitioners specify nodes, edges, and navigation rules by hand. This is general but scales poorly. Our \textit{Dimensions API} addresses this.

Our new vocabulary mirrors how practitioners already think about their data, much as a grammar of graphics lets a designer write a visualization at the level of data fields rather than primitive marks. Rather than specifying a graph directly, a practitioner describes the \textit{dimensions} of their data, the meaningful axes along which a user might want to navigate, and a single build step constructs the full node-link structure automatically. Notably, this enables systems to programmatically construct navigable structures as long as dimensions are configurable. Each dimension declares a \texttt{type} (\texttt{categorical} or \texttt{numerical}) and a \texttt{dimensionKey} naming the field it traverses, together with a \texttt{behavior} block that governs traversal (how navigation behaves at group boundaries, and whether leaf nodes are reachable laterally across divisions) and an \texttt{operations} block that governs how the data is shaped into the hierarchy (sorting, filtering, and binning numerical ranges). \Cref{app:dimensions} gives a deeper look into the API's parameters.

This data-dimensions framing both aligns with and departs from \textit{Olli}~\cite{Blanco2022olli}, which similarly lets authors express a navigable structure over a chart from a higher-level specification rather than hand-built nodes. Both share the premise that fields, not graph primitives, are the right authoring unit. The difference is what the specification produces and where it lives: \textit{Olli} compiles a chart specification into a traversable tree that an author does not subsequently manipulate as a graph, whereas the \textit{Dimensions API} generates an explicit Data Navigator node-link structure whose every node, edge, and rule remains addressable, so that the declarative output is itself the object a practitioner can then inspect in the \textit{Inspector} and edit directly in \textit{Skeleton}.

\section{\textit{Skeleton}: System Design}
\label{sec:skeleton}

With a visual structure renderer (the \textit{Inspector}) and a declarative abstraction for producing navigation structures (the \textit{Dimensions API}), the remaining problem was to bring these capabilities into an integrated, direct-manipulation authoring environment where practitioners could not only see structure but manipulate, test, and iterate on it with immediate feedback. \textit{Skeleton} is that environment. It integrates both and extends them with a guided preparation phase, a spatial rendering canvas, and a live testing mode, reversing the \textit{Inspector}'s code-first direction: practitioners design a navigation structure visually and inspect the code representation as a consequence of that design. Each of its authoring techniques makes visible a specific property of non-visual interaction that sighted practitioners otherwise cannot see during authoring: the topology of what is navigable, the spatial mapping of where navigation lives over the chart, the semantics of what is announced, and the temporal sequence in which a user encounters nodes.

A principle runs through the environment: everything \textit{Skeleton} generates is an editable default, not a fixed output. The \textit{Dimensions API} proposes a topology, the scaffold proposes spatial positions and group outlines, and the preparation wizard proposes labels and rules, but each is materialized as concrete nodes, edges, positions, and text that a practitioner can then select and override by direct manipulation, or bypass entirely by constructing nodes and edges manually over an image. The declarative and automated layers lower the cost of a reasonable starting point; they do not constrain what the final structure can be.

\subsection{Staging: Input and Preparation for Authoring}

The authoring workflow proceeds through four stages: upload, prepare, edit, and test. Making a visualization accessible involves decisions about what is navigable, how navigation is triggered, and where focus indicators appear in space. These decisions are related but distinct, and collapsing them into a single undifferentiated interface, as code-only workflows effectively do, makes each one harder to reason about. The stage architecture surfaces them as separate concerns. It also preserves a direct correspondence between what practitioners see in the interface and the structure of the Data Navigator API: each stage maps onto a distinct layer of the API, lowering the barrier to moving from visual authoring into code when production deployment requires it.

\textbf{Upload.} The upload phase is deliberately permissive. Practitioners can bring a dataset, a chart image, both, or neither. When a dataset is present, \textit{Skeleton} parses its fields and infers dimension types automatically, producing a default, starting configuration for the \textit{Prepare} step. When only an image is present, practitioners proceed directly to editing and construct nodes and edges manually over the image. Because dimensions are derived from whatever fields a dataset actually carries rather than from a fixed, pre-registered schema, the approach does not require the data to be known in advance: a different dataset simply yields a different set of inferred dimensions, and a visualization with no tabular dataset at all is handled through the image-only path. This was motivated by our geologic map co-design work: many bespoke visualizations do not have a single underlying dataset, and any tool that requires structured data as a precondition excludes the cases that need it most. \textit{Skeleton} can be applied to any 2D image surface, not only to visualizations in the conventional sense.

\textbf{Prepare.} The prepare stage addresses a hard authoring bottleneck: not placing nodes, but deciding what structure to build at all. A practitioner who has never designed a navigation structure faces an open configuration space with no obvious entry point. The prep stage presents a four-chapter Q\&A wizard that moves through authoring decisions sequentially: (1) whether the chart should have a root node and what it should announce, (2) which data fields should be navigable dimensions and how those dimensions should behave at their boundaries, (3) which keyboard interactions each dimension should be assigned to, and (4) what text labels each level of the hierarchy should produce. Each chapter is accompanied by an illustrative schematic diagram of which part of the hierarchy is being edited as well as a diagram showing examples of what these decisions look like when showing on a chart. The wizard's output populates a configuration in the editor that practitioners can then inspect, refine, and revise.

\subsection{Edit: Interacting with Topology, Layout, and Semantics}

\begin{figure}[h]
  \centering 
  \includegraphics[width=\columnwidth, alt={A scatterplot, stacked bar, and line chart. Each has a graph next to it that shows nodes and edges. On the other side of each chart is a version of itself with grides, outlines, and marks that represent the various shapes and locations that the nodes and edges take over the chart's space.}]{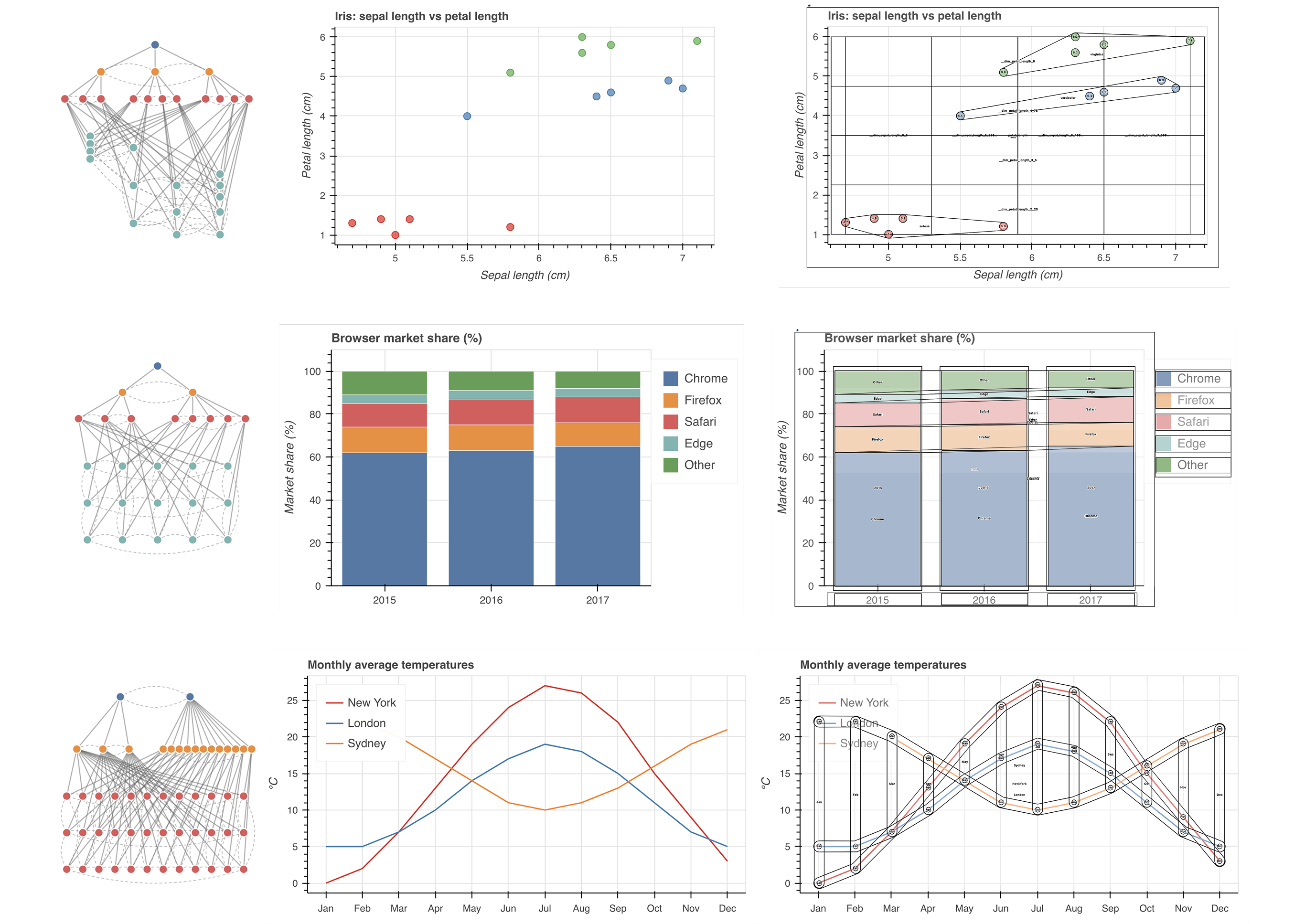}
  \caption{%
  	Input data transformed into a navigable structure using the \textit{Dimensions API} and visualized with our \textit{Inspector} gadget (left). The input chart (middle). The navigable structure is transformed and drawn over the chart using the \textit{Scaffolding Engine} (right).%
  }
  \label{fig:scaffold}
\end{figure}

\textbf{Seeing the system, seeing the experience.} The editor is \textit{Skeleton}'s primary authoring environment (\Cref{fig:hero}) and presents two interlinked representations of the same navigation structure. A schema panel shows the structure as an abstract hierarchical tree layout that makes levels and parent-child relationships immediately readable. A graph canvas shows the same structure rendered as geometric elements positioned over the uploaded chart image, representing what an end user would encounter spatially. These two representations are bidirectionally linked: selecting a node in either view propagates the selection to the other, so practitioners can simultaneously hold in mind both the abstract topology of what is navigable and the spatial instantiation of where that navigation will live. The dual-view design makes visible a real conceptual divide between the system's model of navigation and a user's experience of it, one that practitioners recognize once they can see it, even without prior vocabulary for it.

\textbf{Leveraging visualization as a scaffolding engine.} Manually positioning leaf nodes over each data mark is the most mechanical step in the authoring workflow, and it scales poorly with dataset size. In our early pilot sessions, actually placing nodes in the canvas space was the slowest and most tedious part of the process. To address this, \textit{Skeleton} includes a scaffold tool (\Cref{fig:scaffold}) that automates spatial placement by repurposing Vega~\cite{Satyanarayan2017VegaLite} as a layout engine.

The scaffold renders a Vega chart specification to a hidden, off-screen container and extracts node positions via the library's internal scale and view APIs, using the rendering engine purely as a coordinate computation step: no actual chart is ever shown to the practitioner. Coincidentally, none of our co-designers were crafting visualizations using Vega or Vega-Lite (or derivatives), yet the Vega rendering engine could faithfully reconstruct every necessary mark position over the underlying, inaccessible data visualization provided to \textit{Skeleton}. 

Additionally, positions and outline strategies for category-level group nodes are computed from the leaf positions using geometric algorithms: a union of child node paths, a convex hull, a grid over numerical bounds, or a bounding rectangle. These designs were consistently produced as ideal treatments during co-design, so we chose them for our starting set of outline strategies.

The scaffold is optional and works by generating synthetic placeholder positions that practitioners can adjust manually. It dramatically improved authoring speed: in light pilot tests, a research team member completed a three-dimension structure without scaffolding in 8 minutes 22 seconds and with scaffolding in 56 seconds. A co-designer completed the task incorrectly (failing to account for one dimension's division nodes) in 13 minutes 7 seconds without scaffolding, and correctly in 2 minutes 44 seconds with it.

\textbf{Specifying token patterns, editing instances.} Selecting any node populates a properties panel with spatial properties (position, size, shape) and semantic properties (ARIA role, description, and a label template editor; \Cref{fig:label}). The label template allows practitioners to assemble the text a screen reader will announce at that node from tokens drawn from data fields, including aggregate statistics at group-level nodes and precise data values at leaf nodes~\cite{Jones2024}. Editing labels can apply to all nodes of the same type or to a single instance.

At the bottom of the semantic section, a live preview displays the full assembled announcement string in the exact form a screen reader would produce: role, semantics, group membership, and label combined into a single rendered output that updates in real time. Prior to this preview, understanding what would be announced at a given node required running a screen reader and navigating to it sequentially. In deeply nested structures, this meant spending several seconds listening to labels and drilling in. This was consistently the main point where bugs were produced and missed during our co-design work.

The preview makes text announcements inspectable as visible, editable objects. This surfaces a class of highly specific, low-level problems that code-only workflows leave invisible: redundancies in announced text, missing contextual framing, and label ordering and punctuation that affects comprehension and reading speed. Of all the authoring decisions \textit{Skeleton} exposes, label templates involve the most degrees of freedom and have the most direct bearing on the quality of the non-visual experience. Getting the structure right ensures navigability; getting the labels right determines whether navigation communicates anything meaningful.

\subsection{Test: Debugging Interaction Interactively}
\label{sec:testing}

The testing stage allows practitioners to navigate the structure they have built using the same keyboard input and navigation rules that assistive technologies would use, without leaving the tool. When a practitioner enters testing, the three Data Navigator modules are instantiated in sequence: the structure module rebuilds the navigation graph from the current configuration, the rendering module creates an HTML layer positioned over the chart image at each node's spatial coordinates, and the input module registers keyboard listeners for all navigation rules. The result is a live, keyboard-navigable structure. An event log records navigation events in order, letting practitioners verify that all nodes are reachable and that label sequences make sense when encountered serially rather than read simultaneously.

The abstract graph continues to display during testing. As the practitioner navigates, the focused node is highlighted simultaneously in the canvas (showing its spatial position over the chart image) and in the abstract graph (showing its structural position in the hierarchy). Practitioners can verify at a glance both where focus is and what role it occupies. A mobile-friendly text-chat mode is also available, in which practitioners navigate by typing natural language commands, motivated by the Adobe collaboration (\Cref{sec:designsystem}).



\textbf{Practice-based validation.} The testing stage also served, during development, as the primary debugging interface for \textit{Skeleton}'s own data pipeline. Additionally, CD2, a subject matter expert who professionally evaluates interfaces for screen reader access and is familiar with other visualization navigation systems, used \textit{Skeleton}'s testing stage to evaluate navigation output across several chart types and dimension configurations: line charts (3 configurations), bar charts (2), stacked bar charts (3), and scatterplots (4). This evaluation followed a manual, systematic approach combining standards-based criteria with expert screen reader testing. Scatterplots required the most iteration, surfacing bugs in Data Navigator's core library that were then fixed. CD2 also recommended that we use list-based navigation while in the editor (not in testing) in case users build themselves into a keyboard trap.

\section{User Study}
\label{sec:study}

Our co-design work was motivated by the problems of the communities the research team was embedded in, and our co-designers were deeply familiar with the problem space, having spent months or years working on accessible navigation. We still needed to understand what happens when practitioners who are \textit{not} embedded in this process design, author, and debug navigation structure visually: whether the representations we built are legible to them, whether the techniques change how they reason about accessible design, and what new questions or problems emerge when navigation structure becomes visible. These are empirical questions that required a study.

To evaluate how \textit{Skeleton} influences the way practitioners engage with accessible design, we conducted an in-situ interview study with 8 participants across visualization design, engineering, and research. The study was conducted remotely over video call, took approximately 45--60 minutes per session, and was approved by our IRB (2025\_00000285). Participants gave informed, written consent before our sessions and provided verbal consent at the start of each session. Video and audio were not recorded, however some participants consented to share the data/image they brought to the study as well as screenshots of their work; data collection was note-based throughout.

\subsection{Participants}

Participants were recruited through snowball sampling within the visualization and accessibility community, and through referrals from co-designers involved in our earlier collaborations. We asked each participant to self-report their primary work role (engineering, research, design, or student) and their existing level of accessibility expertise on a 1--5 Likert scale. We also asked whether the visualizations they build are ever bespoke, that is, custom rather than instances of a recognizable, standard chart type. This distinction mattered because bespoke visualizations represent an especially underserved case in accessible design tooling: no library pattern applies, and every navigation structure must be designed from scratch. Participants were not compensated.

\subsection{Procedure}

Each 45-minute session proceeded in four phases. Before the session, all participants were asked to prepare a chart image they were currently working on or had recently built, for use in the third phase.

\textbf{Phase 1: Introduction and demographics (5 minutes).} After obtaining verbal consent and recording a pseudonym, we collected self-reported role, accessibility expertise level, and whether the participant regularly builds bespoke visualizations.


\textbf{Phase 2: Generic chart think-aloud~\cite{Alhadreti2018} (10 minutes).} Participants used \textit{Skeleton} on a provided bar chart and dataset of fruit counts (Apples, Pears, Nectarines, Plums, Grapes), asked to design an accessible navigation experience for a screen reader user with no instructions on how the tool worked. The tool loaded with a default structure having both a categorical dimension (\texttt{fruit}) and a numerical dimension (\texttt{count}) active. This default was intentionally problematic for two reasons: numerical navigation sorts by count value and groups data into subdivided ranges, producing a traversal order different than the visual layout and adds an additional, largely unhelpful level in the hierarchy for such a simple chart. We observed how participants reasoned about what they saw and whether and how they noticed this extra dimension.

\textbf{Phase 3: Own-chart think-aloud (15 minutes).} Participants loaded their own chart image and attempted the same task, except they were also asked to explain their graphic to the research team (purpose, role, data, and domain). This phase was open-ended: charts ranged from standard types to bespoke visualizations, and the goal was to observe how participants reasoned about navigation structure when the context was their own work.

\textbf{Phase 4: Reflective interview (15 minutes).} We conducted a semi-structured interview in which participants reflected on their decisions in Phase 2 versus Phase 3, their experience with the generic versus their own chart, and their assessment of the tool's capabilities and limitations. We asked what felt possible or impossible, what they wanted to do that they could not, and what they found themselves thinking about that they had not considered in Phase 2.

\subsection{Analysis}

Notes from each session were compiled into a shared document. Participant quotes reported in the results are reconstructed from these researcher notes, not verbatim transcripts. We analyzed the data using a combination of thematic analysis~\cite{Braun2006Using} and affinity diagramming~\cite{Harboe2015}, iterating across both methods to surface recurring patterns while preserving the specificity of individual participant experiences. Analysis attended particularly to differences in how participants engaged with accessible design before and after using the tool, the range of input modalities and user scenarios they considered, and moments when participants reconsidered or wished to redesign their own visualizations.

\section{Results}
\label{sec:results}

We organize our findings into five themes that emerged from thematic analysis and affinity diagramming across all eight sessions. Each theme captures a qualitative pattern in how practitioners engaged with accessible navigation design when its structure was made visible and manipulable. We report these findings descriptively and ground them in specific participant moments; interpretation follows in the Discussion.

\subsection{Seeing Navigation Made Structural Problems Legible as Design Problems}
\label{sec:results-legible}

The generic bar chart in Phase 2 loaded with an intentionally problematic default: both a categorical dimension (\texttt{fruit}) and a numerical dimension (\texttt{count}) were active, producing overlapping navigation structures with different traversal orders over the same data. This configuration is a poor design choice, arguably a design failure, but one that would be difficult to detect in code alone~(\ref{rq:R1}, \ref{rq:R4}).

Participants varied widely in how quickly they recognized the problem. P1 turned off the numerical dimension within seconds of seeing the editor, without commenting on it. Most participants, however, initially struggled to understand what they were seeing, remarking on the unfamiliar structure: ``what is this? what are these?'' when encountering the numerical divisions for the first time. P8 spent time trying to guess what the extra divisions represented but did not remove them during Phase 2, only realizing during Phase 3 that the additional dimension was ``probably bad.'' P2 and P5 expressed suspicion early: P5 asked, ``Is this too much data? This seems like way too much to just navigate through,'' and P2 noted, ``I feel like a lot of data points would be bad, yeah? Like, too many at once is bad?''


The testing stage (\Cref{sec:testing}) proved critical for resolution. P4, P5, and P7 each removed the extra dimension only after navigating the structure with keyboard input in testing mode, where the traversal sequence made the redundancy experientially apparent. In total, five of eight participants resolved the problem during the session: P1 and P2 during editing, and P4, P5, and P7 after testing.

Beyond the intentional default, participants identified other problems through visual inspection. P8 reacted to a generated node name: ``Okay dim\_fruit node\ldots that is horrible, what is that?'' During Phase 3, P7 looked at the edges of their multi-line chart and asked, ``are all these bad? Is it bad that I don't even really know what the takeaway of this [structure] is?'' P3, seeing a full hierarchy for their own simple six-item bar chart (during Phase 3), concluded ``I should just skip the root and grouping and go straight to the data. This seems like too many steps.'' In each case, the visual representation of their navigation structure motivated judgment about potential negative design qualities.

\subsection{Practitioners Developed a Designerly Interest in What Constitutes Good Navigation}
\label{sec:results-designerly}

The most pervasive pattern across sessions was that participants began asking design questions about navigation quality, unprompted by any instruction or guidance from the research team~(\ref{rq:R2}). These questions went beyond identifying errors: participants wanted to know what \textit{good} navigation would be for their charts.


P2, working through the bar chart in Phase 2, deliberated over boundary behavior: ``Loop back or stop? I don't think there is a right way. I will just pick \textit{fruit} for now and \textit{loop} and see what this does.'' P6, who brought a bespoke flower visualization, wondered how to translate the affective quality of their chart: ``I think my visualization should be more about the vibes, but I don't know how to make the alt text have good vibes. What is \textit{fun}?'' P4 asked fundamental questions about the interaction model itself: ``Why do screen readers and keyboards have to work this way? Do people like that?\ldots why do we navigate?'' And later after testing, P4 concluded, ``I bet we should make this \textit{faster}'' before cutting out the additional numerical dimension in Phase 2.


Several participants engaged with the concept of narrative and flow. P8 articulated this as a question about the goal of the visualization: ``sometimes I want a big picture, not precision. I may want to drill down a little\ldots'' and observed that ``we think too much in terms of components\ldots sometimes accuracy isn't the actual goal, it's getting a general sense of something.'' P4, upon discovering that \textit{Skeleton} supports text-based input, asked, ``How do you make that good, though? Like chatGPT, or do people want to, like, interact with the chart [elements]?'

During the interview, several participants explicitly requested guidance. P1, P2, P3, and P6 wanted to see examples of well-designed navigation experiences. P1, P3, P4, P5, P6, and P7 wanted embedded guidelines within the tool. P2 and P4 actually used web search to look for ``chart navigation for accessibility guidelines'' (P2) and ``accessible viz screen reader design'' (P4). P3 was interested in automation and heuristics that could suggest reasonable defaults. These requests are consistent with the pattern~(\ref{rq:R2}): practitioners who could see the design space wanted orientation within it.

\subsection{Iteration Was Substantive and Self-directed}
\label{sec:results-iteration}

Every participant iterated on their navigation designs, and this iteration was neither perfunctory nor prompted by the research team~(\ref{rq:R4}). Participants revisited decisions, revised configurations, and in some cases restructured their entire approach after encountering their design in a new stage of the tool.

The most sustained iteration concerned labels. Every participant spent the largest share of their authoring time in the label template editor (\Cref{fig:label}), editing the text that a screen reader would announce at each node. This editing was granular: participants wrote full sentences, rearranged the order of data tokens, debated whether to include field keys alongside values or values alone, experimented with how to name groups and individual elements, and considered the length and density of the resulting announcements. At the division and dimension levels, some participants added multiple aggregate statistics (count, sum, average, range, trend) and then returned to trim them. P2, for instance, edited data point labels, left them, and returned to revise them two additional times: ``This label is way too complicated, I think.''

A second, distinct pattern of iteration emerged around testing. The editor displays all navigation nodes simultaneously, showing the full structure at a glance. The testing stage (\Cref{sec:testing}), by contrast, shows only the currently focused node, highlighting it in the canvas one at a time as the practitioner navigates. This difference consistently prompted participants to revise. Several returned to the editor after testing to adjust group node outlines, because outline strategies that looked distinct when displayed simultaneously (such as bounding rectangles vs. convex hulls) became harder to differentiate when encountered one at a time. Others adjusted their dimension configurations: P3 and P5 restructured their dimensions after testing, and P4, P6, and P8 experimented with different key bindings and navigation rules. P2 used the testing stage specifically to identify labels that needed revision.

The most extensive iteration came from P1, who returned to the preparation wizard after reaching the testing stage and re-took the entire wizard from scratch. Seeing the full structure in testing motivated them to re-evaluate their earliest decisions. They reduced their navigation from three dimensions to one, producing a navigation experience they described as deliberately minimal. In the reflective interview, P1 explained that they had been thinking about trimming excess, joking that they were ``working on a data-to-word ratio.''

The scaffolding tool shaped the pace and character of these iterations. Participants who used the scaffold (\Cref{fig:scaffold}) generally required only minor manual adjustment of node positions, spending one to two minutes refining placements in-aggregate before moving on. Iteration on spatial layout was primarily about the appearance of group-level outlines: because the scaffold computed leaf positions from the chart's encoding, the main remaining spatial decision was how division and dimension nodes should visually indicate grouping.

\subsection{Seeing Navigation Prompted Practitioners to Reconsider the Architecture of Their Own Charts}
\label{sec:results-reconsider}

An unexpected finding was that working with \textit{Skeleton} prompted several participants to question the design of the visualization they had brought, not just the navigation structure they were building over it~(\ref{rq:R4}). Five participants (P4, P5, P6, P7, P8) expressed a desire to simplify their own charts after seeing what the corresponding navigation structure required. Three (P1, P5, P6) considered whether a different chart type might serve the same communicative goal with a simpler navigational architecture. And in 2 cases (P1, P6), participants questioned whether a chart was the right medium at all.

P5, who brought a scatter plot with over two thousand data points, initially tried to build a navigation structure over the full dataset, recognized that the result was unmanageable, and then asked: ``do I even need visualization?'' They went on to wonder whether the information could be communicated as ``a few sentences or like, some data [users] can prompt'' (referring to using a large-language model). P6, who brought a bespoke flower visualization in which each petal encoded a different variable, initially tried to express the chart's full complexity in navigation (grouping by flower and then navigating by petal) but then reversed course: ``okay, what if we actually just treat this like a bar chart?'' Using the scaffold tool, they produced a simple list-style navigation that worked well for their data, and reflected: ``my visualization has a lot, but actually, this could be pretty easy to navigate, I think.'' P6 also wondered whether there was a non-chart way to ``tell their story,'' and in further discussion described something closer to a scrollytelling article or guided walkthrough than a single interactive chart.

\begin{figure}[h]
  \centering 
  \includegraphics[width=\columnwidth, alt={A voronoi pie chart with a large slice emphasized. A large text caption reads, Small assignments: 26 small assignments will make p 45\% of your grade. These exercises, games, quizzes, activites, readings, critiques, and discussions are to be completed in-class on the same day they're issued. Showing up and being present will go a long way to earning you full credit.}]{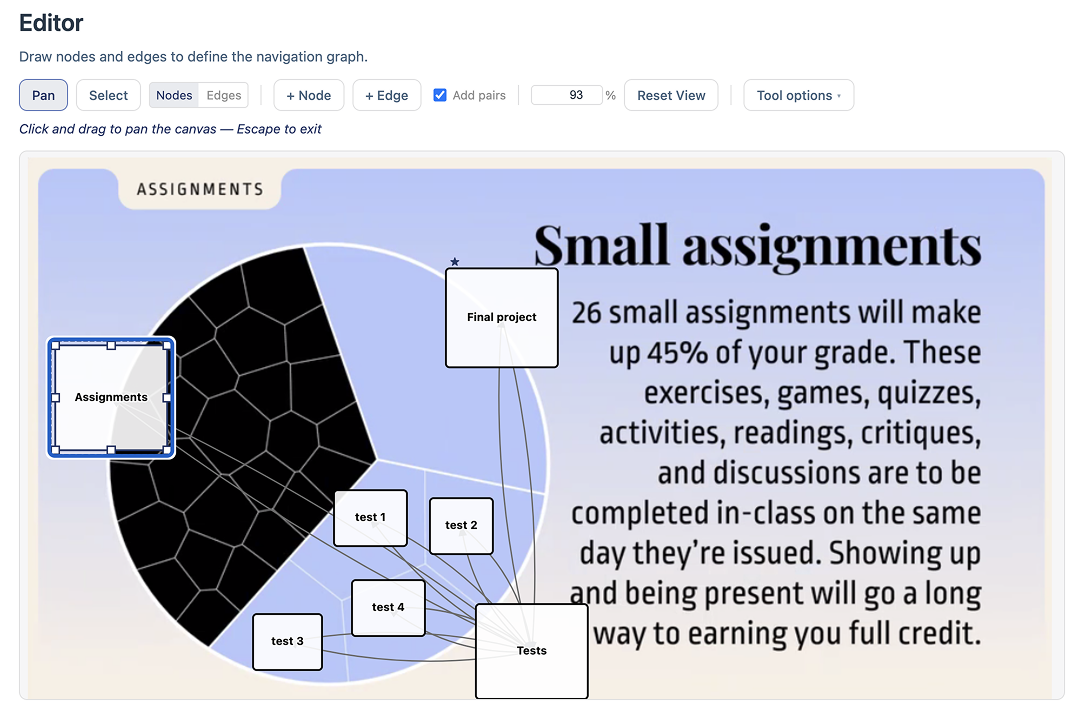}
  \caption{%
  	Re-creation of P8's moment of realization, placing nodes manually: not every element in their voronoi pie chart \textit{should} be navigable.%
  }
  \label{fig:voronoi}
\end{figure}

P8's case was the most striking. They brought a voronoi pie chart (\Cref{fig:voronoi}) used to communicate to students that 26 assignments make up 45\% of their grade, the largest slice of the pie. Working through \textit{Skeleton}, P8 first considered making all 26 voronoi cells navigable, then considered making only the three main slices of the pie navigable, and then questioned whether navigation was needed at all. The ``point of this,'' as they described it, was to communicate a single ratio; students did not need to traverse individual cells. P8 concluded that well-written alternative text was sufficient for the screen reader experience and chose not to build any structure. They also reflected on the design of the visualization itself, concluding that the voronoi treatment served a visual purpose (it is visually striking) that was separable from the informational purpose (communicating a grade breakdown). They kept the visual design and simplified the non-visual experience accordingly.

\subsection{Experiencing Keyboard Navigation Surfaced a Broader Range of Users and Input Technologies}
\label{sec:results-keyboard}

The testing stage (\Cref{sec:testing}) provided what was, for most participants, their first experience of keyboard navigation through a data visualization. Only three participants (P1, P2, P8) reported prior experience using a screen reader, and only two (P1, P8) had previously navigated a visualization using keyboard input alone. Yet during the study, every participant navigated using a keyboard.

Several participants responded to this experience with genuine engagement. P4 reacted with enthusiasm: ``Woah, this is incredible. Wait, what other visualizations can do this?'' and later, ``Maybe you could make a visualization game, where you can move around the data?'' P7 said, ``I love this. I want to make charts with this.'' P5, who had not previously encountered keyboard-navigable charts, said, ``I didn't know you could do this.''

The experience also expanded participants' sense of who these structures serve~(\ref{rq:R2}, \ref{rq:R4}). P8 adjusted a node's spatial dimensions and said, ``Let's make the hitbox as big as possible here\ldots for my neighbor with Parkinson's to click these,'' treating the navigation structure as relevant to motor accessibility, not only screen reader access. P3, observing the visual focus indicators that appeared during scaffold-based placement, remarked that ``this is for more people than [someone] blind,'' recognizing that sighted keyboard users and users with partial vision also rely on visible focus states. During the interview, P4 asked sustained questions about what kinds of technologies different people with disabilities use, and P4 and P7 both asked why focus indication was designed to be visible rather than invisible.

P4's curiosity extended to alternative input modalities. Upon learning that \textit{Skeleton} supports text-based navigation and (due to leveraging Data Navigator) also supports a wide array of other input modalities, they asked, ``Can you talk at it, too? Is that what some people do?'' and followed up with, ``How cool is that? How do you make that good, though?'' P8 raised a concern about discoverability in the current interaction model: ``I've never used j or w to drill out, always shift arrow or option arrow\ldots'' and worried that a user might not know how to exit a nested level. These observations reflect a broad mental model of the user, from an abstract ``screen reader user'' to a person with multiple capabilities, preferences, and interaction patterns~(\ref{rq:R2}).

\section{Discussion}
\label{sec:discussion}

\subsection{Visibility as a Precondition for Iteration}

The central finding of this work is not that \textit{Skeleton} taught sighted practitioners how to design accessible navigation, but that it gave them something to react to. When navigation structure was invisible, our co-designers would only seek to verify whether navigation followed our design documentation. Once visible, questions shifted to whether it was good, why, how it could be improved, and what else it could do. Visibility encouraged continuous design iteration.

This mechanism is consistent with Sch\"on's account of reflective practice~\cite{Schon1996Relective}: designers iterate by externalizing a representation, perceiving its properties, and responding to what they see. The representation talks back, and the designer adjusts. But this loop requires a representation. In our co-design work, we assumed that the gap was hand-off between design and development, and the slow process of manual verification. Instead, the gap is that development was not participating in, and encouraging, further design reflection. In a developer's code-only workflow, navigation structure has no externalization that supports this kind of perceptual engagement. A practitioner can read the code that specifies a navigation graph, but they cannot see the graph, cannot perceive its topology at a glance, cannot notice that a label is redundant or that a hierarchy is too deep by looking at it. The reflective loop is disrupted and occluded at the first step.

\textit{Skeleton} encourages this loop by rendering navigation structure in a form that supports the same kind of perceptual judgment that visualization practitioners already apply to every other aspect of their work. The results show what happened when this loop was available: every participant iterated, substantively and self-directedly (\Cref{sec:results-iteration}). They revised labels repeatedly, restructured dimensions after testing, adjusted group outlines when sequential traversal revealed problems that simultaneous display had hidden. P1 restarted the entire preparation wizard after seeing their structure in testing mode. These are not the behaviors of practitioners following a specification; they are the behaviors of practitioners negotiating with a design material.

The co-design work (\Cref{sec:codesign}) provides complementary evidence: in each collaboration, sighted practitioners reached for visual representations when reasoning about navigation, through node-link diagrams in Figma, schema sketches in Miro, and annotated wireframes on paper. They were already thinking visually about non-visual structure; the development tooling simply had not caught up. The practical implication is broad: if sighted authors depend on visibility, then any authoring workflow that keeps navigation structure invisible limits design iteration. Making structure visible does not guarantee good design, but it is a precondition for the kind of sustained, judgment-driven refinement that good design requires.


\subsection{From Compliance to Design}

A consistent pattern across the accessibility literature is that practitioners frame accessibility as a compliance problem: a set of requirements to satisfy, a checklist to complete, a legal or institutional obligation to meet~\cite{Joyner2022Wild, Sharif2024Challenges, Elavsky2022Chartability}. The framing matters because compliance and design orient practitioners toward fundamentally different activities. Compliance asks: ``does this pass?'' Design asks: ``is this good?'' Our results suggest that \textit{Skeleton} produced a partial shift from the first orientation to the second. Participants' initial instincts clustered around compliance-oriented responses: provide alternative text, follow guidelines, ask an expert (\Cref{sec:results-designerly}). But alongside that desire for guidance, they began doing something compliance framing does not typically produce: they encountered complexity and then \textit{iterated} (\Cref{sec:results-iteration}). This sustained, self-directed refinement is the behavioral signature of design, not compliance. As P4 put it: ``I'd love to have someone blind actually just with me while I make this, but I also understand that I should learn what makes a good experience too.''

The shift extended beyond the non-visual experience itself. As reported in \Cref{sec:results-reconsider}, five participants reconsidered the design of the visualization they had brought, not just the navigation structure. Making the accessibility consequences of visual design choices visible prompted practitioners to question whether a different chart type, a simpler encoding, or a non-chart medium might better serve their communicative goals. This interrelation between non-visual and visual design suggests that the widespread treatment of accessibility as a compliance activity may be partly a consequence of tooling that offers no legible, manipulable design surface. Auditing frameworks are valuable, but they are \textit{evaluative} tools, not \textit{authoring} tools. The field needs both.

\subsection{Bespoke Visualizations as an Unaddressed Accessibility Research Problem}



Diagrams, infographics, and data-driven illustrations are often one-off, custom designs with bespoke symbols, layouts, and visual languages. These representations are increasingly common in journalism, scientific communication, personal projects, art, and public-facing data work, and they represent the cases where accessibility-focused tools are needed most and available least. \textit{Skeleton}'s image-based workflow (upload any 2D image, place nodes manually) provides a starting point, but the study made clear that bespoke visualizations need more than node placement. They need support for reasoning about what navigational structure (if any) is appropriate when no template applies, a research problem that remains largely unexplored. It is worth being precise about where generality holds and where it stops. Because the \textit{Dimensions API} derives structure from whatever fields a dataset carries rather than from a fixed schema, the data side of the approach already extends to datasets not known in advance; what does not yet generalize is the prior, harder question of which dimensions, if any, a bespoke visual should expose when its visual form was never organized around a tabular structure to begin with. Generalizing the abstraction is therefore not primarily a matter of supporting more schemas but of supporting the design judgment that precedes choosing a schema at all.

\subsection{What Visualization Owes Accessibility}



Our approach, to make visual non-visual experiences, should not be limited to data visualization's own accessibility challenges. Navigation structure is a foundational component of accessible experience across domains: PDF and document reading order, web page structures, and software application layouts. In each of these areas, sighted practitioners author non-visual experiences without visual feedback, and in each, the same gap between design intent and verifiable outcome constrains quality. Testing is slow, error prone, and requires expertise in assistive technology use. Visual tooling for authoring, inspecting, and debugging non-visual structure~(\ref{rq:R3}) is a tractable and high-value problem across application domains.

There is also a deeper question about what visibility can accomplish in principle. Work on multi-modal authoring environments has argued for de-centering visual representation and treating modalities as equal partners in the design process~\cite{Zong2024}, an important commitment. \textit{Skeleton} holds this in tension: it re-centers visual representation as the medium through which sighted practitioners engage with non-visual structure. We believe this is justified pragmatically, because sighted authors need to articulate navigation design in their own perceptual language before they can reason about it, and this paper provides evidence that they do. But articulating a design in one's own language is not the same as understanding how it will be experienced in someone else's. The risk we want to name is that making non-visual structure visible to sighted practitioners could be mistaken for making it \textit{understood}, when in fact it makes it \textit{designable}. The fuller design process requires collaboration with disabled users, not as an occasional supplement but as a regular practice. \textit{Skeleton} can make that collaboration more productive by giving both parties a shared representation or space of translation between representations, but it cannot replace it.



What visualization owes accessibility, then, is not simply better output but authoring tools that better stimulate reasoning, both individually and collaborative, about design.



\section{Limitations and Future Work}

\textit{Skeleton} surfaces design questions but does not answer them: several participants asked what constitutes good navigation, and the tool had nothing authoritative to offer. As \Cref{sec:discussion} argues, our study evaluated whether making structure visible stimulated design consideration, not whether the resulting designs were good for the people who will use them, which is a distinct question that remains open. CD2's expert screen reader evaluation of \textit{Skeleton}'s navigation output (\Cref{sec:testing}) provided practice-based validation for several common chart types and surfaced concrete bugs, but this evaluation was neither comprehensive nor controlled: many configurations remain untested, and expert review is not a substitute for evaluation with a broader population of end users.


Our approach using \textit{Skeleton} as a design probe only with sighted participants is a limitation and, we believe, the right sequencing: \textit{Skeleton} improves the intentionality and iterability of what sighted practitioners produce, which is a necessary precondition for a subsequent evaluation or future collaborative design work between sighted and blind authors. Further research and evaluation should close this loop, ideally to engage how mixed-ability teams co-design multi-modal data experiences.





A further boundary concerns the kind of visualization \textit{Skeleton} targets. Highly interactive or dynamic visualizations, where selecting an element reloads the data, where marks animate, or where the navigable structure itself changes in response to user interaction, are not yet addressed.  We see this as an important future direction rather than a fundamental obstacle: the same argument for making structure visible applies with more force when there are several structures to coordinate, and an inspectable representation may be precisely what makes a dynamic navigation experience tractable to author.

\textit{Skeleton} is also a prototype with substantial work remaining. Not within the scope of the paper, but our participants provided ample feedback on the functionality of the prototype itself. The most urgent gap is export functionality: practitioners can design and inspect navigation structures in the tool but cannot yet produce deployable output.



\section{Conclusion}

Accessible navigation structure has long occupied an awkward position in visualization practice: known to matter, difficult to design, and invisible to the people responsible for building it. Without a way to see what they were making, sighted practitioners could not catch errors, could not iterate, and could not develop the kind of considered judgment that good design requires; accessibility remained downstream of every other decision not because of any single failure, but because the authoring conditions did not support anything else. \textit{Skeleton} demonstrates that those conditions can be changed. Making navigation structure visible and manipulable (as an interactive graph rendered over the spatial layout of a real visualization, with live label previews and testable traversal) shifted how practitioners engaged with and reasoned about accessible design, prompting them to ask whether the design of their structure was \textit{good} rather than merely whether it passed.

The implication generalizes beyond our tool: wherever a non-visual structure or user experience is authored without a visible representation, building one the author can react to and manipulate invites them to design. What the paper leaves open is more than what it closes. We used \textit{Skeleton} as a design probe with sighted practitioners; we did not evaluate the structures they produced with the end users those structures are meant to serve, and the broader disciplinary conversation about what visualization research owes accessibility, and what methods might transfer between them, has more questions than answers. We offer \textit{Skeleton} not as a solution to these problems but as evidence that engaging them directly, with the full weight of visualization's methodological tradition, is both possible and fruitful.

\acknowledgments{%
	This work was supported by a grant from Apple, Inc. Any views, opinions, findings, and conclusions or recommendations expressed in this material are those of the authors and should not be interpreted as reflecting the views, policies or position, either expressed or implied, of Apple Inc.

    The in-situ co-design and co-engineering work we performed was made possible for our partners by a CZI EOSS Cycle 6 Grant, a research gift from Adobe, Inc., and a small grant from the University of Wisconsin. Any views, opinions, findings, and conclusions or recommendations expressed in this material are those of the authors and should not be interpreted as reflecting the views, policies or position, either expressed or implied, of these organizations.

    I want to especially thank our fantastic, brilliant co-designers and collaborators, who made this work possible.
}

\bibliographystyle{abbrv-doi-hyperref}

\bibliography{skeleton}

@article{Lo,
  author = {L. Y. Lo and A. Gupta and K. Shigyo and A. Wu and E. Bertini and H. Qu},
  journal = {Computer Graphics Forum},
  month = jun,
  number = {3},
  pages = {515--525},
  title = {Misinformed by Visualization: What Do We Learn From Misinformative Visualizations?},
  volume = {41},
  year = {2022}
}

@article{Elavsky2023,
  author = {Elavsky, Frank and Nadolskis, Lucas and Moritz, Dominik},
  doi = {10.1109/tvcg.2023.3327393},
  issn = {2160-9306},
  journal = {IEEE TVCG},
  publisher = {Institute of Electrical and Electronics Engineers (IEEE)},
  title = {Data Navigator: An Accessibility-Centered Data Navigation Toolkit},
  url = {https://doi.org/10.1109/tvcg.2023.3327393},
  year = {2023}
}

@inproceedings{Thompson2023Chart,
  author = {Thompson, John R and Martinez, Jesse J and Sarikaya, Alper and Cutrell, Edward and Lee, Bongshin},
  booktitle = {Proceedings of the 2023 CHI Conference on Human Factors in Computing Systems},
  doi = {10.1145/3544548.3581186},
  month = apr,
  pages = {1–18},
  publisher = {ACM},
  series = {CHI ’23},
  title = {Chart Reader: Accessible Visualization Experiences Designed with Screen Reader Users},
  url = {https://doi.org/10.1145/3544548.3581186},
  year = {2023}
}

@inproceedings{Sharif2022VoxLens,
  articleno = {478},
  author = {Sharif, Ather and Wang, Olivia H. and Muongchan, Alida T. and Reinecke, Katharina and Wobbrock, Jacob O.},
  booktitle = {CHI Conference on Human Factors in Computing Systems},
  doi = {10.1145/3491102.3517431},
  month = apr,
  numpages = {19},
  pages = {1–19},
  publisher = {ACM},
  series = {CHI ’22},
  title = {VoxLens: Making Online Data Visualizations Accessible with an Interactive JavaScript Plug-In},
  url = {https://doi.org/10.1145/3491102.3517431},
  year = {2022}
}

@article{Farahani2023,
  author = {Farahani, Ali Mazraeh and Adibi, Peyman and Ehsani, Mohammad Saeed and Hutter, Hans-Peter and Darvishy, Alireza},
  doi = {10.1109/access.2023.3298050},
  issn = {2169-3536},
  journal = {IEEE Access},
  pages = {76202–76221},
  publisher = {Institute of Electrical and Electronics Engineers (IEEE)},
  title = {Automatic Chart Understanding: A Review},
  url = {https://doi.org/10.1109/access.2023.3298050},
  volume = {11},
  year = {2023}
}

@inproceedings{Kim2023Exploring,
  author = {Kim, Jiho and Srinivasan, Arjun and Kim, Nam Wook and Kim, Yea-Seul},
  booktitle = {Proceedings of the 2023 CHI Conference on Human Factors in Computing Systems},
  doi = {10.1145/3544548.3581532},
  month = apr,
  pages = {1–15},
  publisher = {ACM},
  series = {CHI ’23},
  title = {Exploring Chart Question Answering for Blind and Low Vision Users},
  url = {https://doi.org/10.1145/3544548.3581532},
  year = {2023}
}

@inproceedings{Jones2024,
  author = {Jones, Shuli and Pedraza Pineros, Isabella and Hajas, Daniel and Zong, Jonathan and Satyanarayan, Arvind},
  booktitle = {Proceedings of the CHI Conference on Human Factors in Computing Systems},
  collection = {CHI ’24},
  doi = {10.1145/3613904.3641970},
  month = may,
  pages = {1–14},
  publisher = {ACM},
  series = {CHI ’24},
  title = {“Customization is Key”: Reconfigurable Textual Tokens for Accessible Data Visualizations},
  url = {https://doi.org/10.1145/3613904.3641970},
  year = {2024}
}

@book{Hermann2011,
  address = {Berlin, Germany},
  editor = {Hermann, Thomas and Hunt, Andy and Neuhoff, John G},
  language = {en},
  month = dec,
  note = {ISBN: 978-3-8325-2819-5},
  isbn = {9783832528195},
  publisher = {Logos Verlag Berlin},
  title = {The Sonification Handbook},
  year = {2011}
}

@inproceedings{Kim2024,
  author = {Kim, Hyeok and Kim, Yea-Seul and Hullman, Jessica},
  booktitle = {Proceedings of the CHI Conference on Human Factors in Computing Systems},
  collection = {CHI ’24},
  doi = {10.1145/3613904.3642442},
  month = may,
  pages = {1–19},
  publisher = {ACM},
  series = {CHI ’24},
  title = {Erie: A Declarative Grammar for Data Sonification},
  url = {https://doi.org/10.1145/3613904.3642442},
  year = {2024}
}

@inproceedings{Zong2024,
  author = {Zong, Jonathan and Pedraza Pineros, Isabella and Chen, Mengzhu (Katie) and Hajas, Daniel and Satyanarayan, Arvind},
  booktitle = {Proceedings of the CHI Conference on Human Factors in Computing Systems},
  collection = {CHI ’24},
  doi = {10.1145/3613904.3641996},
  month = may,
  pages = {1–20},
  publisher = {ACM},
  series = {CHI ’24},
  title = {Umwelt: Accessible Structured Editing of Multi-Modal Data Representations},
  url = {https://doi.org/10.1145/3613904.3641996},
  year = {2024}
}

@inproceedings{Race2023,
  author = {Race, Lauren and Fleet, Chancey and Montour, Danielle and Yazzolino, Lindsay and Salsiccia, Marco and Ferrari, Claire and Wells-Jensen, Sheri and Hurst, Amy},
  booktitle = {The 25th International ACM SIGACCESS Conference on Computers and Accessibility},
  collection = {ASSETS ’23},
  doi = {10.1145/3597638.3614546},
  month = oct,
  pages = {1–8},
  publisher = {ACM},
  series = {ASSETS ’23},
  title = {Designing While Blind: Nonvisual Tools and Inclusive Workflows for Tactile Graphic Creation},
  url = {https://doi.org/10.1145/3597638.3614546},
  year = {2023}
}

@inproceedings{Seo2024,
  author = {Seo, JooYoung and Xia, Yilin and Lee, Bongshin and Mccurry, Sean and Yam, Yu Jun},
  booktitle = {Proceedings of the CHI Conference on Human Factors in Computing Systems},
  collection = {CHI ’24},
  doi = {10.1145/3613904.3642730},
  month = may,
  pages = {1–22},
  publisher = {ACM},
  series = {CHI ’24},
  title = {MAIDR: Making Statistical Visualizations Accessible with Multimodal Data Representation},
  url = {https://doi.org/10.1145/3613904.3642730},
  year = {2024}
}

@article{Reinders2025,
  author = {Reinders, Samuel and Butler, Matthew and Zukerman, Ingrid and Lee, Bongshin and Qu, Lizhen and Marriott, Kim},
  doi = {10.1109/tvcg.2024.3456358},
  issn = {2160-9306},
  journal = {IEEE TVCG},
  month = jan,
  number = {1},
  pages = {864–874},
  publisher = {Institute of Electrical and Electronics Engineers (IEEE)},
  title = {When Refreshable Tactile Displays Meet Conversational Agents: Investigating Accessible Data Presentation and Analysis with Touch and Speech},
  url = {https://doi.org/10.1109/tvcg.2024.3456358},
  volume = {31},
  year = {2025}
}

@inproceedings{Heer2007,
  author = {Heer, Jeffrey and Agrawala, Maneesh},
  booktitle = {2007 IEEE Symposium on Visual Analytics Science and Technology},
  doi = {10.1109/vast.2007.4389011},
  pages = {171–178},
  publisher = {IEEE},
  title = {Design Considerations for Collaborative Visual Analytics},
  url = {https://doi.org/10.1109/vast.2007.4389011},
  year = {2007}
}

@article{Liu2014,
  author = {Liu, Zhicheng and Heer, Jeffrey},
  doi = {10.1109/tvcg.2014.2346452},
  issn = {1077-2626},
  journal = {IEEE TVCG},
  month = dec,
  number = {12},
  pages = {2122–2131},
  publisher = {Institute of Electrical and Electronics Engineers (IEEE)},
  title = {The Effects of Interactive Latency on Exploratory Visual Analysis},
  url = {https://doi.org/10.1109/tvcg.2014.2346452},
  volume = {20},
  year = {2014}
}

@inproceedings{Shneiderman1992,
  author = {Shneiderman, Ben and Williamson, Christopher and Ahlberg, Christopher},
  booktitle = {Proceedings of the SIGCHI conference on Human factors in computing systems},
  collection = {CHI ’92},
  doi = {10.1145/142750.143082},
  pages = {669–670},
  publisher = {ACM Press},
  series = {CHI ’92},
  title = {Dynamic queries: database searching by direct manipulation},
  url = {https://doi.org/10.1145/142750.143082},
  year = {1992}
}

@article{Endert2013,
  author = {Endert, A. and Bradel, L. and North, C.},
  doi = {10.1109/mcg.2013.53},
  issn = {0272-1716},
  journal = {IEEE Computer Graphics and Applications},
  month = jul,
  number = {4},
  pages = {6-13},
  publisher = {Institute of Electrical and Electronics Engineers (IEEE)},
  title = {Beyond Control Panels: Direct Manipulation for Visual Analytics},
  url = {https://doi.org/10.1109/mcg.2013.53},
  volume = {33},
  year = {2013}
}

@inproceedings{Alhadreti2018,
  author = {Alhadreti, Obead and Mayhew, Pam},
  booktitle = {Proceedings of the 2018 CHI Conference on Human Factors in Computing Systems},
  collection = {CHI ’18},
  doi = {10.1145/3173574.3173618},
  month = apr,
  pages = {1–12},
  publisher = {ACM},
  series = {CHI ’18},
  title = {Rethinking Thinking Aloud: A Comparison of Three Think-Aloud Protocols},
  url = {https://doi.org/10.1145/3173574.3173618},
  year = {2018}
}

@inproceedings{Harboe2015,
  author = {Harboe, Gunnar and Huang, Elaine M.},
  booktitle = {Proceedings of the 33rd Annual ACM Conference on Human Factors in Computing Systems},
  collection = {CHI ’15},
  doi = {10.1145/2702123.2702561},
  month = apr,
  pages = {95–104},
  publisher = {ACM},
  series = {CHI ’15},
  title = {Real-World Affinity Diagramming Practices: Bridging the Paper-Digital Gap},
  url = {https://doi.org/10.1145/2702123.2702561},
  year = {2015}
}

@inproceedings{Fisher2012,
  author = {Fisher, Danyel and Popov, Igor and Drucker, Steven and schraefel, m.c.},
  booktitle = {Proceedings of the SIGCHI Conference on Human Factors in Computing Systems},
  collection = {CHI ’12},
  doi = {10.1145/2207676.2208294},
  month = may,
  pages = {1673–1682},
  publisher = {ACM},
  series = {CHI ’12},
  title = {Trust me, i’m partially right: incremental visualization lets analysts explore large datasets faster},
  url = {https://doi.org/10.1145/2207676.2208294},
  year = {2012}
}

@article{Marriott,
  author = {Marriott, Kim and Lee, Bongshin and Butler, Matthew and Cutrell, Ed and Ellis, Kirsten and Goncu, Cagatay and Hearst, Marti and McCoy, Kathleen and Szafir, Danielle Albers},
  doi = {10.1145/3457875},
  issn = {1558-3449},
  journal = {Interactions},
  month = apr,
  number = {3},
  pages = {47–51},
  publisher = {Association for Computing Machinery (ACM)},
  title = {Inclusive data visualization for people with disabilities: a call to action},
  url = {https://doi.org/10.1145/3457875},
  volume = {28},
  year = {2021}
}

@inproceedings{Butler2021Technology,
  author = {Butler, Matthew and Holloway, Leona M and Reinders, Samuel and Goncu, Cagatay and Marriott, Kim},
  booktitle = {Proceedings of the 2021 CHI Conference on Human Factors in Computing Systems},
  collection = {CHI ’21},
  doi = {10.1145/3411764.3445207},
  month = may,
  pages = {1–15},
  publisher = {ACM},
  series = {CHI ’21},
  title = {Technology Developments in Touch-Based Accessible Graphics: A Systematic Review of Research 2010-2020},
  url = {https://doi.org/10.1145/3411764.3445207},
  year = {2021}
}

@inproceedings{Holloway2022,
  author = {Holloway, Leona M and Goncu, Cagatay and Ilsar, Alon and Butler, Matthew and Marriott, Kim},
  booktitle = {CHI Conference on Human Factors in Computing Systems},
  collection = {CHI ’22},
  doi = {10.1145/3491102.3517465},
  month = apr,
  pages = {1–13},
  publisher = {ACM},
  series = {CHI ’22},
  title = {Infosonics: Accessible Infographics for People who are Blind using Sonification and Voice},
  url = {https://doi.org/10.1145/3491102.3517465},
  year = {2022}
}

@misc{Holloway2024,
  author = {Holloway,  Leona and Cracknell,  Peter and Stephens,  Kate and Fanshawe,  Melissa and Reinders,  Samuel and Marriott,  Kim and Butler,  Matthew},
  copyright = {arXiv.org perpetual,  non-exclusive license},
  doi = {10.48550/ARXIV.2401.15836},
  publisher = {arXiv},
  title = {Refreshable Tactile Displays for Accessible Data Visualisation},
  url = {https://doi.org/10.48550/arxiv.2401.15836},
  year = {2024}
}

@article{Marriott2026,
  author = {Marriott, Kim and Butler, Matthew and Holloway, Leona and Jolley, William and Lee, Bongshin and Maguire, Bruce and Szafir, Danielle Albers},
  doi = {10.1109/tvcg.2025.3634254},
  issn = {2160-9306},
  journal = {IEEE TVCG},
  month = jan,
  number = {1},
  pages = {659–669},
  publisher = {Institute of Electrical and Electronics Engineers (IEEE)},
  title = {From Vision to Touch: Bridging Visual and Tactile Principles for Accessible Data Representation},
  url = {https://doi.org/10.1109/tvcg.2025.3634254},
  volume = {32},
  year = {2026}
}

@inproceedings{Sharif2021,
  author = {Sharif, Ather and Chintalapati, Sanjana Shivani and Wobbrock, Jacob O. and Reinecke, Katharina},
  booktitle = {Proceedings of the 23rd International ACM SIGACCESS Conference on Computers and Accessibility},
  collection = {ASSETS ’21},
  doi = {10.1145/3441852.3471202},
  month = oct,
  pages = {1–16},
  publisher = {ACM},
  series = {ASSETS ’21},
  title = {Understanding Screen-Reader Users’ Experiences with Online Data Visualizations},
  url = {https://doi.org/10.1145/3441852.3471202},
  year = {2021}
}

@inproceedings{Mei2025,
  author = {Mei⁎, Catherine and Pollock⁎, Josh and Hajas, Daniel and Zong, Jonathan and Satyanarayan, Arvind},
  bib = {===========},
  booktitle = {Proceedings of the 27th International ACM SIGACCESS Conference on Computers and Accessibility},
  collection = {ASSETS ’25},
  doi = {10.1145/3663547.3746342},
  month = oct,
  pages = {1–17},
  publisher = {ACM},
  series = {ASSETS ’25},
  title = {Benthic: Perceptually Congruent Structures for Accessible Charts and Diagrams},
  url = {https://doi.org/10.1145/3663547.3746342},
  year = {2025}
}

@inproceedings{Blanco2022olli,
  author = {Matt Blanco AND Jonathan Zong AND Arvind Satyanarayan},
  booktitle = {IEEE VIS Posters},
  note = {\url{https://vis.csail.mit.edu/pubs/olli}},
  title = {{Olli: An Extensible Visualization Library for Screen Reader Accessibility}},
  url = {https://vis.csail.mit.edu/pubs/olli},
  year = {2022}
}

@inproceedings{Bornschein2015Collaborative,
  address = {New York, NY, USA},
  author = {Bornschein, Jens and Prescher, Denise and Weber, Gerhard},
  booktitle = {Proceedings of the 17th International ACM SIGACCESS Conference on Computers \& Accessibility - ASSETS ’15},
  doi = {10.1145/2700648.2809869},
  month = oct,
  organization = {Association for Computing Machinery},
  pages = {117–126},
  publisher = {ACM Press},
  series = {ASSETS \textquoteright{}15},
  title = {Collaborative {Creation} of {Digital} {Tactile} {Graphics}},
  url = {https://doi.org/10.1145/2700648.2809869},
  year = {2015}
}

@article{Bostock2011D3,
  address = {USA},
  author = {Bostock, M. and Ogievetsky, V. and Heer, J.},
  doi = {10.1109/tvcg.2011.185},
  issn = {1077-2626},
  issue_date = {December 2011},
  journal = {IEEE TVCG},
  month = dec,
  number = {12},
  numpages = {9},
  pages = {2301–2309},
  publisher = {Institute of Electrical and Electronics Engineers (IEEE)},
  title = {D³ Data-Driven Documents},
  url = {https://doi.org/10.1109/tvcg.2011.185},
  volume = {17},
  year = {2011}
}

@article{Braun2006Using,
  author = {Braun, Virginia and Clarke, Victoria},
  doi = {10.1191/1478088706qp063oa},
  issn = {1478-0895},
  journal = {Qualitative Research in Psychology},
  month = jan,
  number = {2},
  pages = {77–101},
  publisher = {Informa UK Limited},
  title = {Using thematic analysis in psychology},
  url = {https://doi.org/10.1191/1478088706qp063oa},
  volume = {3},
  year = {2006}
}

@article{Brewster2002Visualization,
  author = {Brewster, S.},
  doi = {10.1080/09638280110111388},
  issn = {1464-5165},
  journal = {Disability and Rehabilitation},
  month = jan,
  number = {11-12},
  pages = {613–621},
  publisher = {Informa UK Limited},
  title = {Visualization tools for blind people using multiple modalities},
  url = {https://doi.org/10.1080/09638280110111388},
  volume = {24},
  year = {2002}
}

@article{Chundury2022Towards,
  author = {Chundury, Pramod and Patnaik, Biswaksen and Reyazuddin, Yasmin and Tang, Christine and Lazar, Jonathan and Elmqvist, Niklas},
  doi = {10.1109/tvcg.2021.3114829},
  issn = {2160-9306},
  journal = {IEEE TVCG},
  month = jan,
  number = {1},
  pages = {1084–1094},
  publisher = {Institute of Electrical and Electronics Engineers (IEEE)},
  title = {Towards Understanding Sensory Substitution for Accessible Visualization: An Interview Study},
  url = {https://doi.org/10.1109/tvcg.2021.3114829},
  volume = {28},
  year = {2022}
}

@inproceedings{Cullen2019Co,
  author = {Cullen, Clare and Metatla, Oussama},
  booktitle = {IDC '19},
  collection = {IDC ’19},
  doi = {10.1145/3311927.3323146},
  month = jun,
  pages = {361-373},
  publisher = {ACM},
  series = {IDC ’19},
  title = {Co-designing Inclusive Multisensory Story Mapping with Children with Mixed Visual Abilities},
  url = {https://doi.org/10.1145/3311927.3323146},
  year = {2019}
}

@inproceedings{deGreef2021Interdependent,
  articleno = {36},
  author = {de Greef, Lilian and Moritz, Dominik and Bennett, Cynthia},
  booktitle = {Proceedings of the 23rd International ACM SIGACCESS Conference on Computers and Accessibility},
  doi = {10.1145/3441852.3476468},
  month = oct,
  numpages = {6},
  pages = {1–6},
  publisher = {ACM},
  series = {ASSETS ’21},
  title = {Interdependent Variables: Remotely Designing Tactile Graphics for an Accessible Workflow},
  url = {https://doi.org/10.1145/3441852.3476468},
  year = {2021}
}

@article{Elavsky2022Chartability,
  author = {Elavsky, Frank and Bennett, Cynthia and Moritz, Dominik},
  doi = {10.1111/cgf.14522},
  journal = {Computer Graphics Forum},
  month = jun,
  number = {3},
  pages = {57-70},
  publisher = {Wiley},
  title = {How accessible is my visualization? Evaluating visualization accessibility with Chartability},
  url = {https://doi.org/10.1111/cgf.14522},
  volume = {41},
  year = {2022}
}

@article{Fan2023Accessibility,
  author = {Fan, Danyang and Fay Siu, Alexa and Rao, Hrishikesh and Kim, Gene Sung-Ho and Vazquez, Xavier and Greco, Lucy and O'Modhrain, Sile and Follmer, Sean},
  doi = {10.1145/3557899},
  issn = {1936-7236},
  journal = {ACM Transactions on Accessible Computing},
  month = mar,
  number = {1},
  pages = {1–29},
  publisher = {Association for Computing Machinery (ACM)},
  title = {The Accessibility of Data Visualizations on the Web for Screen Reader Users: Practices and Experiences During COVID-19},
  url = {https://doi.org/10.1145/3557899},
  volume = {16},
  year = {2023}
}

@article{Jung2022Communicating,
  author = {Jung, Crescentia and Mehta, Shubham and Kulkarni, Atharva and Zhao, Yuhang and Kim, Yea-Seul},
  doi = {10.1109/tvcg.2021.3114846},
  issn = {2160-9306},
  journal = {IEEE TVCG},
  month = jan,
  number = {1},
  pages = {1095–1105},
  publisher = {Institute of Electrical and Electronics Engineers (IEEE)},
  title = {Communicating Visualizations without Visuals: Investigation of Visualization Alternative Text for People with Visual Impairments},
  url = {https://doi.org/10.1109/tvcg.2021.3114846},
  volume = {28},
  year = {2022}
}

@inproceedings{Kim2023Explain,
  author = {Kim, Gyeongri and Kim, Jiho and Kim, Yea-Seul},
  booktitle = {Proceedings of the 2023 CHI Conference on Human Factors in Computing Systems},
  collection = {CHI ’23},
  doi = {10.1145/3544548.3581139},
  month = apr,
  pages = {1–13},
  publisher = {ACM},
  series = {CHI ’23},
  title = {“Explain What a Treemap is”: Exploratory Investigation of Strategies for Explaining Unfamiliar Chart to Blind and Low Vision Users},
  url = {https://doi.org/10.1145/3544548.3581139},
  year = {2023}
}

@article{Kim2021Accessible,
  author = {Kim, N. W. and Joyner, S. C. and Riegelhuth, A. and Kim, Y.},
  doi = {10.1111/cgf.14298},
  journal = {Computer Graphics Forum},
  month = jun,
  number = {3},
  pages = {173-188},
  publisher = {Wiley},
  title = {Accessible Visualization: Design Space, Opportunities, and Challenges},
  url = {https://doi.org/10.1111/cgf.14298},
  volume = {40},
  year = {2021}
}

@inproceedings{Lundgard2019Sociotechnical,
  author = {Lundgard, Alan and Lee, Crystal and Satyanarayan, Arvind},
  booktitle = {2019 IEEE Visualization Conference (VIS)},
  doi = {10.1109/visual.2019.8933762},
  month = oct,
  pages = {16–20},
  publisher = {IEEE},
  title = {Sociotechnical Considerations for Accessible Visualization Design},
  url = {https://doi.org/10.1109/visual.2019.8933762},
  year = {2019}
}

@article{Lundgard2022Accessible,
  author = {Lundgard, Alan and Satyanarayan, Arvind},
  doi = {10.1109/tvcg.2021.3114770},
  issn = {2160-9306},
  journal = {IEEE TVCG},
  month = jan,
  number = {1},
  pages = {1073–1083},
  publisher = {Institute of Electrical and Electronics Engineers (IEEE)},
  title = {Accessible Visualization via Natural Language Descriptions: A Four-Level Model of Semantic Content},
  url = {https://doi.org/10.1109/tvcg.2021.3114770},
  volume = {28},
  year = {2022}
}

@article{Mansur1985Sound,
  author = {Mansur, Douglass L. and Blattner, Merra M. and Joy, Kenneth I.},
  doi = {10.1007/bf00996201},
  issn = {1573-689X},
  journal = {Journal of Medical Systems},
  month = jun,
  number = {3},
  pages = {163-174},
  publisher = {Springer Science and Business Media LLC},
  title = {Sound graphs: A numerical data analysis method for the blind},
  url = {https://doi.org/10.1007/bf00996201},
  volume = {9},
  year = {1985}
}

@article{Satyanarayan2017VegaLite,
  author = {Satyanarayan, Arvind and Moritz, Dominik and Wongsuphasawat, Kanit and Heer, Jeffrey},
  doi = {10.1109/tvcg.2016.2599030},
  issn = {1077-2626},
  journal = {IEEE TVCG},
  month = jan,
  number = {1},
  pages = {341–350},
  publisher = {Institute of Electrical and Electronics Engineers (IEEE)},
  title = {Vega-Lite: A Grammar of Interactive Graphics},
  url = {https://doi.org/10.1109/tvcg.2016.2599030},
  volume = {23},
  year = {2017}
}

@misc{Schon1996Relective,
  author = {Schön, Donald and Bennett, John},
  doi = {10.1145/229868.230044},
  isbn = {0201854910},
  journal = {Bringing design to software},
  month = apr,
  pages = {171–189},
  publisher = {ACM},
  title = {Reflective conversation with materials},
  url = {https://doi.org/10.1145/229868.230044},
  year = {1996}
}

@inproceedings{Sharif2018evoGraphs,
  author = {Sharif, Ather and Forouraghi, Babak},
  booktitle = {2018 15th IEEE Annual Consumer Communications \& Networking Conference (CCNC)},
  doi = {10.1109/ccnc.2018.8319239},
  month = jan,
  pages = {1–4},
  publisher = {IEEE},
  title = {evoGraphs — A jQuery plugin to create web accessible graphs},
  url = {https://doi.org/10.1109/ccnc.2018.8319239},
  year = {2018}
}

@incollection{Sorge2016Polyfilling,
  author = {Sorge, Volker},
  booktitle = {Lecture Notes in Computer Science},
  doi = {10.1007/978-3-319-41264-1_6},
  isbn = {9783319412641},
  issn = {1611-3349},
  journal = {Lecture Notes in Computer Science},
  pages = {43–50},
  publisher = {Springer International Publishing},
  title = {Polyfilling Accessible Chemistry Diagrams},
  url = {https://doi.org/10.1007/978-3-319-41264-1_6},
  year = {2016}
}

@techreport{WAI2017Keyboard,
  author = {{WAI}},
  institution = {W3C},
  note = {Accessed: 2026-03-20, \url{https://www.w3.org/WAI/WCAG21/Understanding/keyboard.html}},
  title = {Understanding Success Criterion {2.1.1:} Keyboard},
  type = {{WCAG} Standard},
  url = {https://www.w3.org/WAI/WCAG21/Understanding/keyboard.html},
  year = {2017}
}

@article{Zong2022Rich,
  author = {Zong, Jonathan and Lee, Crystal and Lundgard, Alan and Jang, JiWoong and Hajas, Daniel and Satyanarayan, Arvind},
  doi = {10.1111/cgf.14519},
  journal = {Computer Graphics Forum},
  month = jun,
  number = {3},
  pages = {15-27},
  paper = {===========},
  publisher = {Wiley},
  title = {Rich Screen Reader Experiences for Accessible Data Visualization},
  url = {https://doi.org/10.1111/cgf.14519},
  volume = {41},
  year = {2022}
}

@inproceedings{Joyner2022Wild,
  author = {Joyner, Shakila Cherise S and Riegelhuth, Amalia and Garrity, Kathleen and Kim, Yea-Seul and Kim, Nam Wook},
  booktitle = {CHI Conference on Human Factors in Computing Systems},
  doi = {10.1145/3491102.3517630},
  month = apr,
  pages = {1–19},
  publisher = {ACM},
  title = {Visualization Accessibility in the Wild: Challenges Faced by Visualization Designers},
  url = {https://doi.org/10.1145/3491102.3517630},
  year = {2022}
}

@article{Kim2023Beyond,
  author = {Kim, N. W. and Ataguba, G. and Joyner, S. C. and Zhao, Chuangdian and Im, Hyejin},
  doi = {10.1111/cgf.14833},
  issn = {1467-8659},
  journal = {Computer Graphics Forum},
  month = jun,
  number = {3},
  pages = {323-335},
  publisher = {Wiley},
  title = {Beyond Alternative Text and tables: Comparative Analysis of Visualization Tools and Accessibility Methods},
  url = {https://doi.org/10.1111/cgf.14833},
  volume = {42},
  year = {2023}
}

@inproceedings{Sharif2024Challenges,
  author = {Sharif, Ather and Kim, Joo Gyeong and Xu, Jessie Zijia and Wobbrock, Jacob O.},
  booktitle = {The 26th International ACM SIGACCESS Conference on Computers and Accessibility},
  collection = {ASSETS '24},
  doi = {10.1145/3663548.3675625},
  month = oct,
  pages = {1–20},
  publisher = {ACM},
  series = {ASSETS '24},
  title = {Understanding and Reducing the Challenges Faced by Creators of Accessible Online Data Visualizations},
  url = {https://doi.org/10.1145/3663548.3675625},
  year = {2024}
}

@article{Hayes2011ActionResearch,
  author = {Hayes, Gillian R.},
  doi = {10.1145/1993060.1993065},
  issn = {1073-0516},
  journal = {ACM Transactions on Computer-Human Interaction},
  month = jul,
  number = {3},
  pages = {1-20},
  publisher = {Association for Computing Machinery (ACM)},
  title = {The relationship of action research to human-computer interaction},
  url = {https://doi.org/10.1145/1993060.1993065},
  volume = {18},
  year = {2011}
}

@article{Shneiderman1983DirectManip,
  author = {Shneiderman},
  doi = {10.1109/mc.1983.1654471},
  issn = {0018-9162},
  journal = {Computer},
  month = aug,
  number = {8},
  pages = {57–69},
  publisher = {Institute of Electrical and Electronics Engineers (IEEE)},
  title = {Direct Manipulation: A Step Beyond Programming Languages},
  url = {https://doi.org/10.1109/mc.1983.1654471},
  volume = {16},
  year = {1983}
}

@article{Hutchins1985DirectManip,
  author = {Hutchins, Edwin L. and Hollan, James D. and Norman, Donald A.},
  doi = {10.1207/s15327051hci0104_2},
  issn = {1532-7051},
  journal = {Human–Computer Interaction},
  month = dec,
  number = {4},
  pages = {311–338},
  publisher = {Informa UK Limited},
  title = {Direct Manipulation Interfaces},
  url = {https://doi.org/10.1207/s15327051hci0104_2},
  volume = {1},
  year = {1985}
}

@inproceedings{Mowar2026iTagPDF,
  author = {Mowar, Peya and Steinfeld, Aaron and Bigham, Jeffrey P.},
  booktitle = {Proceedings of the 2026 CHI Conference on Human Factors in Computing Systems},
  doi = {10.1145/3772318.3790289},
  month = apr,
  note = {Best Paper Award},
  pages = {1–17},
  publisher = {ACM},
  title = {{iTagPDF}: Towards Finally Automating {PDF} Accessibility},
  url = {https://doi.org/10.1145/3772318.3790289},
  year = {2026}
}

@inproceedings{Biggs2022Audiom,
  author = {Biggs, Brandon and Toth, Christopher and Stockman, Tony and Coughlan, James M. and Walker, Bruce N.},
  booktitle = {Proceedings of the 27th International Conference on Auditory Display (ICAD 2022)},
  doi = {10.21785/icad2022.027},
  month = jun,
  note = {PMCID: PMC10010675},
  pages = {82-90},
  publisher = {International Community for Auditory Display},
  title = {Evaluation of a Non-Visual Auditory Choropleth and Travel Map Viewer},
  url = {https://doi.org/10.21785/icad2022.027},
  year = {2022}
}

@misc{Eswaramoorthy2025BokehAudit,
  author = {Eswaramoorthy, Pavithra and Elavsky, Frank and Allard, Tania and gabalafou},
  doi = {10.5281/zenodo.14923642},
  month = feb,
  publisher = {Zenodo},
  title = {{Quansight-Labs/bokeh-a11y-audit}},
  url = {https://doi.org/10.5281/zenodo.14923642},
  year = {2025}
}

@inproceedings{Hammad2024GameAware,
  author = {Hammad, Noor and Elavsky, Frank and Moharana, Sanika and Chen, Jessie and Lee, Seyoung and Carrington, Patrick and Moritz, Dominik and Hammer, Jessica and Harpstead, Erik},
  booktitle = {The 26th International ACM SIGACCESS Conference on Computers and Accessibility},
  collection = {ASSETS '24},
  doi = {10.1145/3663548.3675665},
  month = oct,
  pages = {1–13},
  publisher = {ACM},
  series = {ASSETS '24},
  title = {Exploring The Affordances of Game-Aware Streaming to Support Blind and Low Vision Viewers: A Design Probe Study},
  url = {https://doi.org/10.1145/3663548.3675665},
  year = {2024}
}

@inproceedings{Hutchinson2003TechnologyProbes,
  author = {Hutchinson, Hilary and Mackay, Wendy and Westerlund, Bo and Bederson, Benjamin B. and Druin, Allison and Plaisant, Catherine and Beaudouin-Lafon, Michel and Conversy, Stéphane and Evans, Helen and Hansen, Heiko and Roussel, Nicolas and Eiderbäck, Björn},
  booktitle = {Proceedings of the SIGCHI Conference on Human Factors in Computing Systems},
  doi = {10.1145/642611.642616},
  month = apr,
  pages = {17–24},
  publisher = {ACM},
  title = {Technology probes: inspiring design for and with families},
  url = {https://doi.org/10.1145/642611.642616},
  year = {2003}
}

@inproceedings{Zong2025,
  author = {Zong, Jonathan},
  booktitle = {2025 IEEE Workshop on Accessible Data Visualization (AccessViz)},
  doi = {10.1109/accessviz68666.2025.00011},
  month = nov,
  pages = {30–33},
  publisher = {IEEE},
  title = {Using Real Names of Disabled Participant-Contributors to Practice Citational Justice in Accessibility},
  url = {https://doi.org/10.1109/accessviz68666.2025.00011},
  year = {2025}
}

@incollection{Wilkinson2011,
  author = {Wilkinson, Leland},
  booktitle = {Handbook of Computational Statistics},
  doi = {10.1007/978-3-642-21551-3_13},
  isbn = {9783642215513},
  journal = {Handbook of Computational Statistics},
  month = dec,
  pages = {375–414},
  publisher = {Springer Berlin Heidelberg},
  title = {The Grammar of Graphics},
  url = {https://doi.org/10.1007/978-3-642-21551-3_13},
  year = {2011}
}

@article{McNutt2022,
  author = {McNutt, Andrew M.},
  doi = {10.1109/tvcg.2022.3209460},
  issn = {2160-9306},
  journal = {IEEE TVCG},
  pages = {1–11},
  publisher = {Institute of Electrical and Electronics Engineers (IEEE)},
  title = {No Grammar to Rule Them All: A Survey of JSON-style DSLs for Visualization},
  url = {https://doi.org/10.1109/tvcg.2022.3209460},
  year = {2022}
}

@article{ZongAnimated2023,
  author = {Zong, Jonathan and Pollock, Josh and Wootton, Dylan and Satyanarayan, Arvind},
  doi = {10.1109/tvcg.2022.3209369},
  issn = {2160-9306},
  journal = {IEEE TVCG},
  month = jan,
  number = {1},
  pages = {149–159},
  publisher = {Institute of Electrical and Electronics Engineers (IEEE)},
  title = {Animated Vega-Lite: Unifying Animation with a Grammar of Interactive Graphics},
  url = {https://doi.org/10.1109/tvcg.2022.3209369},
  volume = {29},
  year = {2023}
}

@article{Wickham2010,
  author = {Wickham, Hadley},
  doi = {10.1198/jcgs.2009.07098},
  issn = {1537-2715},
  journal = {Journal of Computational and Graphical Statistics},
  month = jan,
  number = {1},
  pages = {3–28},
  publisher = {Informa UK Limited},
  title = {A Layered Grammar of Graphics},
  url = {https://doi.org/10.1198/jcgs.2009.07098},
  volume = {19},
  year = {2010}
}

@inproceedings{Elavsky2025,
  author = {Elavsky, Frank and Bearfield, Cindy Xiong},
  booktitle = {2025 IEEE Workshop on Accessible Data Visualization (AccessViz)},
  doi = {10.1109/accessviz68666.2025.00008},
  month = nov,
  pages = {14–24},
  publisher = {IEEE},
  title = {Playing telephone with generative models},
  url = {https://doi.org/10.1109/accessviz68666.2025.00008},
  year = {2025}
}

@misc{Wimer2026,
  author = {Wimer,  Brianna L. and Kanchi,  Ritesh and Frierson,  Kaija and Potluri,  Venkatesh and Metoyer,  Ronald and Mankoff,  Jennifer and Natsuhara,  Miya and Wang,  Matt X.},
  copyright = {Creative Commons Attribution 4.0 International},
  doi = {10.48550/ARXIV.2601.19168},
  publisher = {arXiv},
  title = {Nonvisual Support for Understanding and Reasoning about Data Structures},
  url = {https://doi.org/10.48550/arxiv.2601.19168},
  year = {2026}
}

@techreport{Rawling2025,
  author = {Rawling, J.E. and Carson, E.C. and Attig, J.W. and Mickelson, D.M. and Mode, W.N. and Johnson, M.D. and Syverson, K.M.},
  doi = {10.54915/xqpw9883},
  institution = {Wisconsin Geological and Natural History Survey},
  publisher = {Wisconsin Geological and Natural History Survey},
  title = {Quaternary Geology of Wisconsin},
  url = {https://doi.org/10.54915/xqpw9883},
  year = {2025}
}

@inproceedings{Sutherland1963Sketchpad,
  address = {New York, NY, USA},
  author = {Sutherland, Ivan E.},
  booktitle = {Proceedings of the May 21-23, 1963, spring joint computer conference on - AFIPS ’63 (Spring)},
  doi = {10.1145/1461551.1461591},
  pages = {329},
  publisher = {ACM Press},
  series = {AFIPS ’63 (Spring)},
  title = {Sketchpad: A Man-Machine Graphical Communication System},
  url = {https://doi.org/10.1145/1461551.1461591},
  year = {1963}
}

@book{Smith1977Pygmalion,
  author = {Smith, David Canfield},
  doi = {10.1007/978-3-0348-5744-4},
  isbn = {9783034857444},
  title = {Pygmalion: A Computer Program to Model and Stimulate Creative Thought},
  url = {https://doi.org/10.1007/978-3-0348-5744-4},
  year = {1977}
}

@article{Teitelbaum1981Cornell,
  author = {Teitelbaum, Tim and Reps, Thomas},
  doi = {10.1145/358746.358755},
  journal = {Communications of the ACM},
  number = {9},
  pages = {563-573},
  publisher = {Association for Computing Machinery (ACM)},
  title = {Cornell program synthesizer},
  url = {https://doi.org/10.1145/358746.358755},
  volume = {24},
  year = {1981}
}

@article{Myers1990Taxonomies,
  author = {Myers, Brad A.},
  doi = {10.1016/S1045-926X(05)80036-9},
  issn = {1045-926X},
  journal = {Journal of Visual Languages \& Computing},
  month = mar,
  number = {1},
  pages = {97–123},
  publisher = {Elsevier BV},
  title = {Taxonomies of Visual Programming and Program Visualization},
  url = {https://doi.org/10.1016/S1045-926X(05)80036-9},
  volume = {1},
  year = {1990}
}

@inproceedings{Thatcher1994,
  series = {Assets ’94},
  title = {Screen reader/2: access to OS/2 and the graphical user interface},
  url = {http://dx.doi.org/10.1145/191028.191039},
  DOI = {10.1145/191028.191039},
  booktitle = {Proceedings of the first annual ACM conference on Assistive technologies  - Assets ’94},
  publisher = {ACM Press},
  author = {Thatcher,  J.},
  year = {1994},
  pages = {39–46},
  collection = {Assets ’94}
}

@inproceedings{Savva2011,
  series = {UIST ’11},
  title = {ReVision: automated classification,  analysis and redesign of chart images},
  url = {http://dx.doi.org/10.1145/2047196.2047247},
  DOI = {10.1145/2047196.2047247},
  booktitle = {Proceedings of the 24th annual ACM symposium on User interface software and technology},
  publisher = {ACM},
  author = {Savva,  Manolis and Kong,  Nicholas and Chhajta,  Arti and Fei-Fei,  Li and Agrawala,  Maneesh and Heer,  Jeffrey},
  year = {2011},
  month = Oct,
  pages = {393–402},
  collection = {UIST ’11}
}

@article{Poco2017,
  title = {Reverse‐Engineering Visualizations: Recovering Visual Encodings from Chart Images},
  volume = {36},
  ISSN = {1467-8659},
  url = {http://dx.doi.org/10.1111/cgf.13193},
  DOI = {10.1111/cgf.13193},
  number = {3},
  journal = {Computer Graphics Forum},
  publisher = {Wiley},
  author = {Poco,  Jorge and Heer,  Jeffrey},
  year = {2017},
  month = June,
  pages = {353–363}
}

\newpage
\appendix

\section{Dimensions API Reference}
\label{app:dimensions}

Each dimension in the \textit{Dimensions API} (\Cref{sec:infrastructure}) declares a \texttt{type} (\texttt{categorical} or \texttt{numerical}, inferred from the data when omitted) and a \texttt{dimensionKey} naming the data field it traverses. Two further groups of properties govern behavior. A \texttt{behavior} block controls traversal: \texttt{extents} sets boundary behavior (\texttt{terminal} stops at the edges of a group, \texttt{circular} wraps around, and \texttt{bridgedCousins} carries focus across to the nearest element of an adjacent group rather than stopping or wrapping), and \texttt{childmostNavigation} determines whether leaf-level nodes are reachable laterally across a dimension's divisions (\texttt{across}) or only within a single division before returning to a parent (\texttt{within}, the default). An \texttt{operations} block controls how the data is shaped into the hierarchy: a \texttt{sortFunction} orders divisions and leaves, a \texttt{filterFunction} selects which data participate, and \texttt{createNumericalSubdivisions} bins a numerical dimension into a chosen number of ranges.

The generated structure is a multi-level hierarchy: each dimension produces a root node, below which division nodes group the data, below which leaf nodes represent individual data points. Multiple dimensions over the same dataset share leaf nodes, so users navigating via different dimensions reach the same data through different paths.

To make the input and output concrete, the examples below use this small four-row dataset of average temperatures, with an explicit \texttt{id} on each row:

\begin{verbatim}
const data = [
  { id: '_0', month: 'Jan', city: 'NYC', temp: 0 },
  { id: '_1', month: 'Jul', city: 'NYC', temp: 25 },
  { id: '_2', month: 'Jan', city: 'Sydney', temp: 22 },
  { id: '_3', month: 'Jul', city: 'Sydney', temp: 8 }
];
\end{verbatim}

\noindent With the Dimensions API, bar chart navigation that would otherwise require constructing every node and edge by hand is expressed as a single dimension declaration:

\begin{verbatim}
dimensions: {
  values: [
    { 
      dimensionKey: 'month',
      type: 'categorical',
      behavior: { extents: 'circular' }
    }
  ]
}
\end{verbatim}

\noindent From this declaration, the build step does three things automatically. It assembles the nodes and the parent-child and sibling edges that connect them, including, when multiple dimensions are declared, the cross-dimension edges that let a user move between hierarchies over shared leaves. It applies the boundary and childmost rules to decide what happens at each edge of the structure. And it assigns keyboard navigation rules to each dimension, drawing from a pool of directional key pairs; because that pool is finite, the API recommends keeping a structure to six dimensions or fewer and asks for explicit navigation rules beyond that point.

This bounds the structures the abstraction expresses by default: it generates the regular, dimensionally-organized hierarchies that recur across common chart types, and structures that do not decompose into a small set of data dimensions (or that exceed the directional-key budget) fall back to explicit graph construction or manual editing in the editor.

The abstraction is otherwise chart-type-agnostic: bar charts, scatter plots, line charts, and layered charts all use the same vocabulary, with different combinations producing different navigation topologies. This directly addressed Bokeh's problem, where no native chart types exist to anchor a pattern: the API mirrors data fields and encoding choices as a set of dimensions, so a thin wrapper can map an arbitrary glyph assembly onto a handful of recurring dimensional shapes (\Cref{sec:opensource}).

The vocabulary composes to richer topologies simply by adding dimensions. Declaring two categorical dimensions over the same dataset, for instance a temperature series recorded by month and by city, produces a dual hierarchy in which left and right traverse one axis and up and down traverse the other, with both sharing the same leaf nodes; here the month dimension wraps at its boundaries while the city dimension stops:

\begin{verbatim}
dimensions: {
  values: [
    {
      dimensionKey: 'month',
      type: 'categorical',
      behavior: {
        extents: 'circular',
        childmostNavigation: 'across'
      }
    },
    {
      dimensionKey: 'city',
      type: 'categorical',
      behavior: {
        extents: 'terminal',
        childmostNavigation: 'across'
      }
    }
  ]
}
\end{verbatim}

\noindent This two-dimension declaration assigns a keyboard control to each axis. \Cref{tab:keys} lists the controls Data Navigator generates for it; the arrow keys move between siblings along each dimension, \texttt{Enter} drills toward the data, and each dimension receives its own drill-out key from the default key pool (here \texttt{W} and \texttt{J}).

\begin{table}[h]
  \centering
  \caption{Keyboard controls generated for the two-dimension (\texttt{month}, \texttt{city}) structure over the sample dataset. Movement along \texttt{month} wraps (\texttt{circular}); movement along \texttt{city} stops at the ends (\texttt{terminal}).}
  \label{tab:keys}
  \begin{tabular}{@{}ll@{}}
    \toprule
    Key & Action \\
    \midrule
    \texttt{$\leftarrow$} / \texttt{$\rightarrow$} & Move to previous / next \texttt{month} (wraps Dec--Jan) \\
    \texttt{$\uparrow$} / \texttt{$\downarrow$} & Move to previous / next \texttt{city} (stops at ends) \\
    \texttt{Enter} & Drill in (dimension $\rightarrow$ division $\rightarrow$ data) \\
    \texttt{W} & Drill out along the \texttt{month} grouping \\
    \texttt{J} & Drill out along the \texttt{city} grouping \\
    \texttt{Esc} & Exit the structure \\
    \bottomrule
  \end{tabular}
\end{table}

\noindent Finally, to show what the build step emits, the two-dimension declaration above, applied to the four-row dataset, expands into the nodes and edges below. Each dimension produces a dimension node carrying its \texttt{dimensionKey}, a set of division nodes (one per distinct value, each tagged with the field it was \texttt{derived} from), and the shared leaf nodes that carry the original rows. A node's \texttt{edges} list the edges it participates in; each edge names a \texttt{source}, a \texttt{target}, and the \texttt{navigationRules} that traverse it. Crucially, a single edge is bidirectional: the \texttt{left}/\texttt{right} pair on one edge means pressing \texttt{left} moves toward its \texttt{source} and \texttt{right} toward its \texttt{target}, so the \texttt{month} and \texttt{city} axes at the leaf level each collapse to one edge with two rules (abbreviated here for space):

\begin{verbatim}
// nodes (excerpt: one division + one leaf per axis)
{
  month: {
    id: 'month',
    data: { dimensionKey: 'month' },
    edges: ['month-Jan', 'any-exit']
  },
  city: {
    id: 'city',
    data: { dimensionKey: 'city' },
    edges: ['city-NYC', 'any-exit']
  },
  'Jan': {
    id: 'Jan',
    derivedNode: 'month',
    data: { month: 'Jan' },
    edges: [
      'Jan-Jul',
      'month-Jan',
      'Jan-_0'
    ]
  },
  'NYC':  {
    id: 'NYC',
    derivedNode: 'city',
    data: { city: 'NYC' },
    edges: [
      'NYC-Sydney',
      'city-NYC',
      'NYC-_0'
    ]
  },
  _0: {
    id: '_0',
    data: { month: 'Jan', city: 'NYC', temp: 0 },
    edges: [
      'leaf-_0-_1',    // month axis (l/r)
      'leaf-_0-_2',    // city axis  (u/d)
      'Jan-_0', 'NYC-_0',
      'any-exit'
    ]
  }
  // Jul, Sydney, and _1/_2/_3
  // follow the same pattern
}

// edges
{
  'month-Jan': {
    source: 'month',
    target: 'Jan',
    navigationRules: ['drill-out_month', 'drill-in']
  },
  'city-NYC': {
    source: 'city',
    target: 'NYC',
    navigationRules: ['drill-out_city', 'drill-in']
  },
  'Jan-Jul': {
    source: 'Jan',
    target: 'Jul',
    navigationRules: ['left','right']
  },
  'NYC-Sydney': {
    source: 'NYC',
    target: 'Sydney',
    navigationRules: ['up','down'] },
  'leaf-_0-_1': {
    source: '_0',
    target: '_1',
    navigationRules: ['left','right']
  },
  'leaf-_0-_2': {
    source: '_0',
    target: '_2',
    navigationRules: ['up','down']
  },
  // ...etc
}

// navigationRules
{
  left: { key: 'ArrowLeft',  direction: 'source' },
  right: { key: 'ArrowRight', direction: 'target' },
  up: { key: 'ArrowUp',    direction: 'source' },
  down: { key: 'ArrowDown',  direction: 'target' },
  'drill-in': { key: 'Enter', direction: 'target' },
  'drill-out_month': {
    key: 'KeyW', 
    direction: 'source'
  },
  'drill-out_city': {
    key: 'KeyJ',
    direction: 'source'
  }
}
\end{verbatim}

\noindent Every node, edge, and rule in this output remains individually addressable: this is the structure the \textit{Inspector} renders and that practitioners select and edit directly in \textit{Skeleton}, rather than a compiled artifact they cannot reopen.

\section{Deterministic Scaffolding, Group Outline Geometries, and Padding Estimation}
\label{app:scaffold}

This appendix details the three technical components behind \textit{Skeleton}'s spatial authoring (\Cref{sec:skeleton}): how the scaffold repurposes a visualization rendering engine to approximate mark positions, how group outlines are computed from those positions, and how a lightweight, non-ML computer vision pass aligns the scaffold to an uploaded chart image. We intentionally chose deterministic math approaches at any opportunity, for speed and inspectability benefits.

\subsection{Vega as a Deterministic Layout Engine}

The scaffold (\Cref{sec:skeleton}, ``Leveraging visualization as a scaffolding engine'') treats a visualization grammar as a coordinate oracle rather than a rendering target. From the active configuration, which carries a chart type, field-to-channel mappings (\texttt{xField}, \texttt{yField}, and an optional \texttt{colorField}), mark parameters, and an outer plot box, \textit{Skeleton} assembles a \textit{Vega-Lite}~\cite{Satyanarayan2017VegaLite} specification and renders it through \texttt{vega-embed} into a hidden, off-screen container using the SVG renderer. The container is never attached to the visible document; it exists only so that \textit{Vega} will compile a full scenegraph and, with it, the scale functions that map data values to pixels.

Positions are then read directly from the compiled view rather than parsed from the rendered output. For each leaf node, \textit{Skeleton} evaluates the view's \texttt{x} and \texttt{y} scales on the corresponding data values: a band scale supplies a bar's horizontal position and, through its \texttt{bandwidth}, its width, while a linear scale maps a quantitative value to a pixel height with the baseline taken at the scale's zero. Stacked bars accumulate per-segment offsets in data space before scaling; clustered bars add the sub-band offset; scatter and line marks evaluate both scales at each point. Because the scale functions are pure, this path needs no DOM measurement and is fully deterministic: identical configurations yield identical coordinates. A secondary path, used when scale evaluation does not apply, parses the hidden SVG and reads mark bounding boxes instead.

The configuration distinguishes the inner mark area from the outer box. \textit{Vega-Lite}'s \texttt{width} and \texttt{height} describe the inner area in which marks are drawn, while a \texttt{padding} object reserves space for axes and labels; \textit{Skeleton} treats \texttt{plotWidth} and \texttt{plotHeight} as the outer, border-box dimensions and derives the inner area by subtracting padding. A mark at normalized position $f$ within the inner area therefore lands at image coordinate $o + p + f\,w_{\mathrm{in}}$, where $o$ is the plot offset, $p$ the leading padding, and $w_{\mathrm{in}}$ the inner extent. This relation is what later makes padding recoverable.

A practical subtlety motivates the rest of this appendix. The engine that produced the uploaded image is generally unknown and is rarely \textit{Vega}; our co-designers authored charts in other tools entirely (\Cref{sec:codesign}). What transfers across engines is the \textit{relative} geometry of the marks, the spacing of bars or the placement of points within the data rectangle, not the absolute pixel offsets, which depend on how much room each particular chart reserved for its axes, ticks, labels, and title. The scaffold reconstructs the relative geometry faithfully; reconciling it with the absolute placement in the image is the task taken up by the padding estimation described below.

\subsection{Computing Group Outlines}

Once leaf positions exist, the group nodes above them (divisions, dimensions, and the root) receive a visible outline computed purely from the geometry of their descendants. Each leaf contributes a rectangle (for bars) or a circle (for points), and a chosen strategy turns the collection into a single SVG path string. These functions live in the Data Navigator core (\texttt{geometry.ts}) and are side-effect free, taking arrays of rectangles or circles and an optional padding term and returning a path \texttt{d} attribute, so they can be reused outside \textit{Skeleton} and packaged for export.

Four strategies cover the cases that recurred during co-design (\Cref{sec:codesign}). A \textit{bounding rectangle} encloses all descendants in the single axis-aligned box given by their combined extent. A \textit{union} emits one rectangular subpath per mark and relies on the SVG fill rule to render disconnected regions, which suits groups whose marks are spatially separated. A \textit{convex hull} collects the corner points of every rectangle (or sixteen sampled points around every circle) and runs a Graham scan, optionally pushing each hull vertex radially outward from the centroid to add padding; this gives a tight, convex boundary appropriate to scatter clusters and bar groups. For line series, an \textit{offset path} traces the polyline of points at a fixed perpendicular distance on each side, using per-vertex averaged normals for miter joins and semicircular arc caps at the ends, producing a closed band that follows the line. Numerical dimensions additionally support a grid drawn over the dimension's value bounds.

Because every outline derives from the same leaf coordinates the scaffold computed, changing a mark position, a chart type, or the padding propagates automatically to the group shapes, and authors edit outline strategy and padding as ordinary node properties rather than drawing shapes by hand.

\subsection{Estimating Plot Padding from the Image}

In this appendix subsection, we specifically outline the final work we did before publication of this project to improve it. In our scaffolding tool, we reduced the time it took for nodes to be placed from eight minutes down to one or less. But human error was still a factor. We wanted to make this process more accurate and \textit{even} faster.

Up to this point, the scaffold's layout is internally consistent but free floating: it knows where marks sit relative to one another, not how the uploaded image split its canvas between the data rectangle and the axis space, which authors originally recovered by manually dragging the four padding values until the scaffold marks settled onto the real marks. We now automate this with a small, fast, dependency-free computer vision pass that runs entirely in the browser, in the spirit of chart reverse-engineering that recovers marks from bitmap images~\cite{Savva2011, Poco2017}.

Detection proceeds on a downscaled copy of the plot region. The image is drawn into an off-screen canvas and a binary foreground mask is built by thresholding the pixel data. The key observation is that data marks are almost always rendered in color while chart chrome (axis lines, gridlines, tick labels, titles, and legend text) is desaturated. A saturation threshold therefore isolates the marks and prevents an axis line from bridging adjacent bars into a single blob; for the rare monochrome chart the pass falls back to a luminance contrast mask. Connected-component labeling groups the mask into candidate marks, each carrying a bounding box, centroid, area, and mean color. Two filters clean the result: components far smaller than the median are discarded as legend swatches, and for line and area charts, where the mark is one long stroke rather than discrete glyphs, the endpoints of the dominant component are used in place of separate marks. The pass returns the extreme detected marks along each axis (leftmost and rightmost, topmost and bottommost) as points in image space.

Padding then follows. A scaffold mark at normalized inner position $f$ sits at $o + p + f\,w_{\mathrm{in}}$, and that $f$ is invariant to padding because it is fixed by the data and scale. Taking the two extreme scaffold marks on an axis, with inner positions $f_{\mathrm{lo}}$ and $f_{\mathrm{hi}}$, and the two detected image positions $d_{\mathrm{lo}}$ and $d_{\mathrm{hi}}$ they should align to, the inner extent and the two padding values follow while the outer box ($o$, $W$) is fixed:
\[
w_{\mathrm{in}}' = \frac{d_{\mathrm{hi}} - d_{\mathrm{lo}}}{f_{\mathrm{hi}} - f_{\mathrm{lo}}}, \quad
p_{\mathrm{lo}}' = d_{\mathrm{lo}} - o - f_{\mathrm{lo}}\,w_{\mathrm{in}}', \quad
p_{\mathrm{hi}}' = W - w_{\mathrm{in}}' - p_{\mathrm{lo}}'.
\]
The two axes are solved independently. Several guards keep the result trustworthy: an axis whose two marks are nearly coincident in $f$ is ill-conditioned and is skipped, negative padding is clamped, and a solved inner extent larger than the outer box signals that the box itself is too small and is reported rather than forced. The pass runs once automatically when an image, a chart type, field mappings, and at least two scaffold marks are all present, and on demand thereafter from a control in the scaffold panel; after applying the padding it re-measures the residual between detected and scaffold marks as a one-shot verification.

\begin{figure}[h]
  \centering 
  \includegraphics[width=\columnwidth, alt={Four visualizations, a bar chart, scatter plot, stacked bar chart, and line chart. Each visualization is shown twice, first with outlines overlaid that are slightly offset and then again with the outlines almost perfectly overlaid on the elements they represent.}]{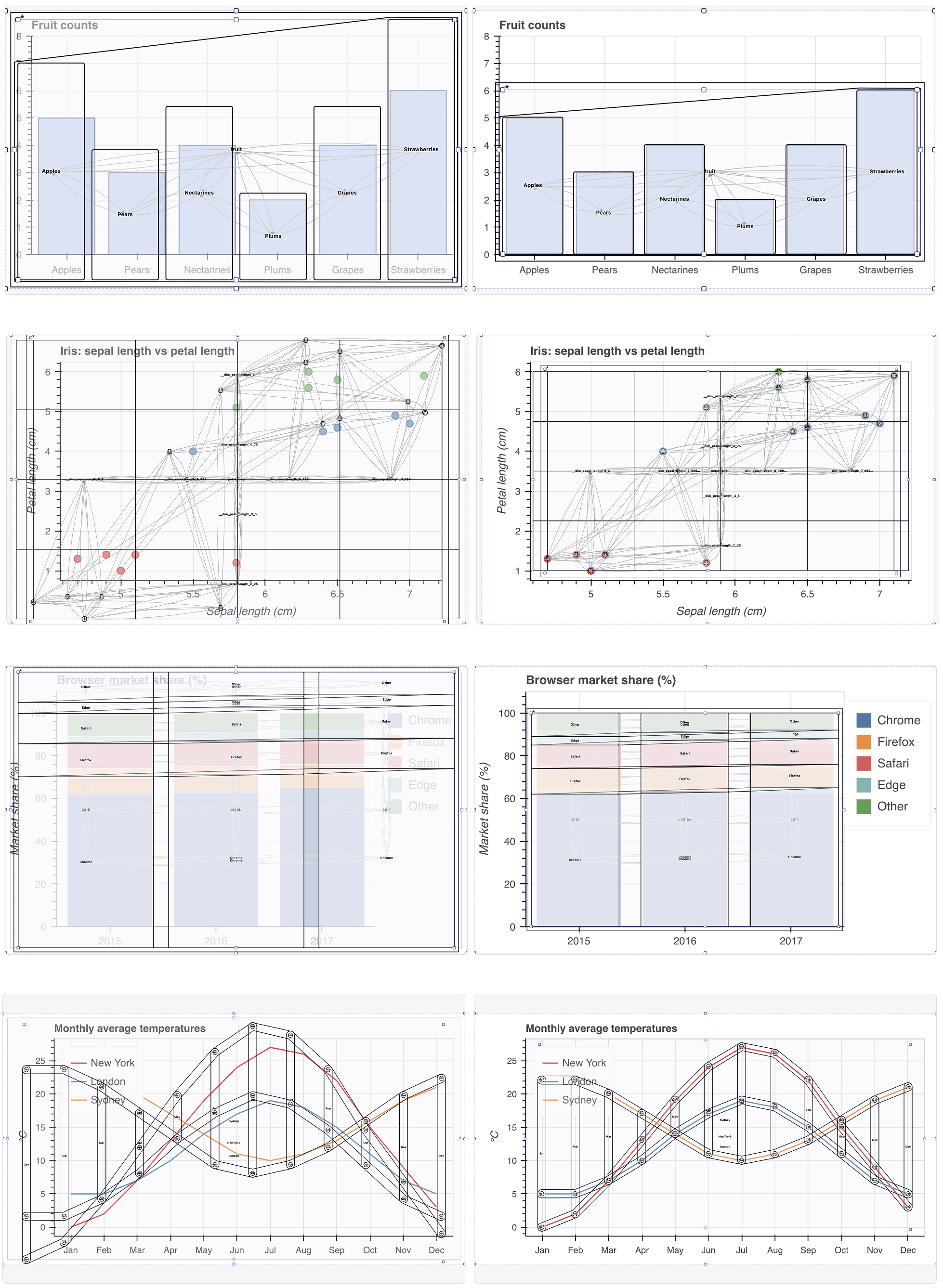}
  \caption{%
  	Computed layout for marks via Vega's visualization engine plus group outlines (left) and with padding estimation using our non-ML computer vision approach (right) for various types of common visualizations.%
  }
  \label{fig:padding}
\end{figure}

In practice this recovers padding consistent with the original chart across the four canonical types we tested (bar, stacked bar, line, and scatter), shown in \Cref{fig:padding}. Rarely is manual adjustment needed, and when it is, it is on the magnitude of one or two pixels of nudging. On the bar example, for instance, the solver recovers a left padding close to the width of the chart's $y$-axis label gutter and a small right padding, with only bar inner padding required for adjustment. The overall time spent by the user is likely on the magnitude of seconds at most, down from a minute. We regard this as promising but early, and a fuller evaluation across more diverse chart images remains future work.

\newpage

\section{Group Label Builder UI}

\begin{figure}[h]
  \centering 
  \includegraphics[width=\columnwidth, alt={Label builder UI. Show interface options, label template filled with generic pattern-building text, and then a label preview shows output "New York, trend for temp: flat, average temp: 13.78, City 3 of 8. Long description of UI details: title "Aggregate summaries from children" unselected count of children, unselected min and max of dropdown with temp selected, unselected sum of dropdown with temp selected, selected average of dropdown with temp selected. "Trend direction and r squared options" x variable dropdown with month selected, y variable dropdown with temp selected, selected trend direction, unselected r squared.}]{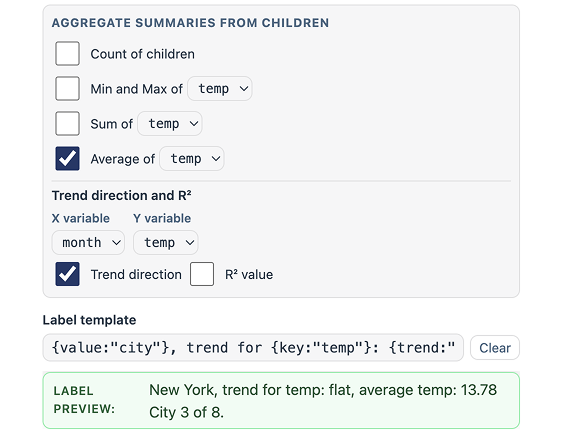}
  \caption{%
  	Group label pattern builder, including an array of aggregate summary options, template formatter field, and preview.%
  }
  \label{fig:label}
\end{figure}

\end{document}